\documentclass[twocolumn,superscriptaddress,floatfix,preprintnumbers,nofootinbib]{revtex4-2}

\usepackage[colorlinks=true,breaklinks=true]{hyperref}
\usepackage[utf8]{inputenc}
\usepackage{longtable}

\hypersetup{urlcolor=Blue,
	    citecolor=Blue,
	    linkcolor=Blue}
\usepackage{url}
\usepackage[normalem]{ulem}
\usepackage{mathtools}
\usepackage[dvipsnames]{xcolor}

\usepackage{orcidlink}
\usepackage{JournalAbbreviations}

\begin{document}

\title{Flavor Equilibration of Supernova Neutrinos: Exploring the Dynamics of Slow Modes}

\author{Ian Padilla-Gay
\orcidlink{0000-0003-2472-3863}} 
\email[Corresponding author: ] {ianpaga@slac.stanford.edu}
\affiliation{Particle Theory Group, SLAC National Accelerator Laboratory, Stanford University, Stanford, CA 94039, USA}

\author{Heng-Hao Chen
\orcidlink{0009-0006-8513-1473}} 
\email{laiho@as.edu.tw}
\affiliation{Institute of Physics, Academia Sinica, Taipei 115201, Taiwan}

\author{Sajad Abbar 
\orcidlink{0000-0001-8276-997X}}
\email{abbar@mpp.mpg.de}
\affiliation{Max-Planck-Institut {f\"ur} Physik (Werner-Heisenberg-Institut), Boltzmannstr. 8, 85748 Garching, Germany}

\author{Meng-Ru Wu
\orcidlink{0000-0003-4960-8706}}
\email{mwu@as.edu.tw}
\affiliation{Institute of Physics, Academia Sinica, Taipei 115201, Taiwan}
\affiliation{Institute of Astronomy and Astrophysics, Academia Sinica, Taipei 106319, Taiwan}
\affiliation{Physics Division, National Center for Theoretical Sciences, Taipei 106319, Taiwan}

\author{Zewei Xiong
\orcidlink{0000-0002-2385-6771}}
\email{z.xiong@gsi.de}
\affiliation{GSI Helmholtzzentrum {f\"ur} Schwerionenforschung, Planckstra{\ss}e 1, D-64291 Darmstadt, Germany}

\preprint{SLAC-PUB-250421}


\begin{abstract}
Neutrinos experience collective flavor conversion in extreme astrophysical environments such as core-collapse supernovae (CCSNe). One manifestation of collective conversion is slow flavor conversion (SFC), which has recently attracted renewed interest owing to its ubiquity across different regions of the supernova environment.
In this study, we systematically examine the evolution of kinematic decoherence in a dense neutrino gas undergoing SFC, considering lepton number asymmetries as large as $30\%$. 
Our findings show that the neutrino gas asymptotically evolves toward a generic state of coarse-grained flavor equilibration which is constrained by approximate lepton number conservation. 
The equilibration occurs within a few factors of the inverse vacuum oscillation frequency, $\omega^{-1}$, which corresponds to (anti)neutrinos reaching near flavor equipartition after a few kilometers for typical supernova neutrino energies. 
Notably, the quasi-steady state of the neutrino number densities can be quantitatively described by the neutrino-antineutrino number density ratio $n_{\bar{\nu}_e}/n_{\nu_e}$ alone. 
Such a simple estimation opens new opportunities for incorporating SFC into CCSN simulations, particularly in regions where SFC develops on scales much shorter than those of collisions. 
\end{abstract}

\maketitle

\section{Introduction}

The fate of explosive astrophysical events such as core-collapse supernovae (CCSNe) depends significantly on accurate modeling of neutrino transport in extreme densities. 
In CCSNe, neutrinos are inevitably produced in vast numbers---roughly $10^{58}$ neutrinos in the MeV energy range---carrying away about 99\% of the gravitational binding energy of the progenitor star. 
Prior to shock formation, the degenerate electron gas is sufficiently energetic to overcome the neutron-proton mass difference $\Delta_{np} = m_n - m_p$, making the reaction $e^- + p \rightarrow n + \nu_e$ favorable, thereby producing a large number of neutrinos. 
Immediately following the bounce, the SN shock forms in neutrino-opaque region, but as it propagates into regions with lower densities within a few milliseconds, the first flash of electron neutrinos—the neutronization burst—is generated. 
As the shock loses energy and stalls, neutrino charged-current interactions with the surrounding matter can re-energize the shock and increase the pressure behind it so that a successful explosion can be launched. 
This constitutes the so-called neutrino heating mechanism, which is currently the most popular scenario of CCSN explosions~\cite{Bethe1985a,Janka2012,Burrows:2020qrp,Janka:2025tvf}.

Given that CCSNe are extremely dense in neutrinos, neutrino flavor conversions (FCs) exhibit unique characteristics in the SN environment. 
In fact, the exceptionally high neutrino densities in CCSNe cause neutrino-neutrino interactions (due to the $Z$ boson exchange) to play a dominant role in their flavor evolution~\cite{pantaleone:1992eq, sigl1993general}. 
Remarkably, this leads to the intriguing, rich, and highly nonlinear phenomenon of \emph{collective neutrino oscillations}, where the entire neutrino gas undergoes FC in a collective manner~\cite{Pastor:2002we, Duan:2005cp, duan:2010bg, Mirizzi:2015eza,volpe2023neutrinos,Johns:2025mlm}.

Understanding neutrino FCs in the CCSN environment is of great importance for several reasons. 
First, FCs affects the expected neutrino signals emerged from CCSNe and needs to be accurately modeled for robustly  interpreting the neutrino signal from the next Galactic CCSN~\cite{Horiuchi2018b,Abbar:2024nhz}. 
Additionally, in CCSNe the nucleosynthesis of heavy elements requires a solid understanding of neutrino-matter interactions, the evolution of the electron number fraction per baryon, $Y_e$, in the neutrino-driven wind, and neutrino-nucleus interactions relevant to the neutrino (induced) nucleosynthesis~\cite{Qian:1996xt,Heger:2003mm,Frohlich:2005ys,Sieverding:2019qet,Ko:2022uqv,Friedland:2023kqp,Wang:2023vkk,Fischer:2023ebq}.
The neutrino flavor composition (and thus FCs) plays a key role in determining $Y_e$ and key rates for neutrino nucleosynthesis, and can modify them through flavor-dependent interactions~\cite{Duan:2010af,Wu:2014kaa,Ko:2019xxm,Xiong:2020ntn,Fujimoto:2022njj,Xiong:2024tac}.
Moreover, neutrino FCs can significantly impact the dynamics of CCSNe and the shock propagation. Schematic implementations of neutrino FCs in dynamical CCSN simulations have demonstrated that including FCs can drastically alter the SN dynamics~\cite{Ehring2023a, Ehring2023b,Mori:2025cke,Wang:2025nii}. 
In fact,  depending on the details of the SN model and the regions where FCs occur most efficiently within the SN environment, FCs can potentially aid or hinder the explosion~\cite{Ehring2023b}.

Over almost the past decade, research on neutrino FCs has focused mainly on understanding fast flavor conversions (FFCs) first discovered in~\cite{Sawyer2005a,Sawyer2009a,Sawyer2016a} (see \cite{Johns:2025mlm} for a recent review and references therein). 
However, the study of neutrino flavor dynamics in neutrino-dense media originally began with what are now called slow flavor conversions (SFCs)~\cite{Pastor:2002we,Duan:2005cp,Hannestad2006g,duan:2006an,duan:2010bg}. 
An important distinction between FFCs and SFCs is that the former can occur in the limit of vanishing vacuum oscillation frequency. Unlike FFCs, which occur only when there is a crossing in the angular distribution of electron neutrinos and antineutrinos, e.g. the electron neutrino lepton number (ELN) angular distribution (assuming heavy-lepton neutrinos and antineutrinos have similar distributions)~\cite{morinaga2022fast,Dasgupta2022a,Dasgupta:2025quc}, SFCs can exist throughout the SN environment for regions outside the proto-neutron star (PNS) after the neutronization burst phase when spectra crossings are expected. 
Despite significant progress in understanding FFCs and their dynamical evolution in realistic CCSN environments~\cite{Nagakura:2023mhr,Nagakura2023e,Xiong:2024tac,Xiong:2024pue,Shalgar:2024gjt,Shalgar:2025oht}, SFCs remain considerably underdeveloped. 
This is mainly because, just as the impact of broken spatial and temporal symmetries on the dynamics of SFCs in SNe began to be explored~\cite{Duan:2014gfa, Abbar:2015fwa, Chakraborty:2016yeg, Abbar:2015mca}, FFCs were discovered. Due to the very short time and length scales of FFCs and their potential presence near neutrino decoupling regions, research in the field quickly shifted toward studying FFCs, leaving SFCs largely overlooked in the literature ever since.

SFCs have recently regained attention, in part due to new insights into the potential impact of FCs in low-density regions ($\rho \sim 10^{9}-10^{10} \ {\rm g/cm^3}$) on SN dynamics~\cite{Ehring2023a, Ehring2023b, Wang:2025nii}. 
For instance, the linear stability analysis based on a spherically symmetric SN model performed in Ref.~\cite{Shalgar:2024gjt} found instabilities associated with SFCs exists at similar densities behind the shock during the accretion phase. 
Refs.~\cite{Fiorillo:2024pns, Fiorillo:2025ank} recently also revisited the linear stability analysis of SFCs and explored different regimes of FCs. 
These analyses quantify that SFC typically exhibit inverse growth rates of the order of order of $\sim \omega^{-1} = 0.7\, \mathrm{km} \times (E/10\, \mathrm{MeV})$, where $\omega = \Delta m^2 / 2E$, $\Delta m^2$ denotes the atmospheric neutrino mass-squared difference, and $E$ represents the neutrino energy. Although the inverse growth scale of SFC is of astrophysical size, the unstable region is typically broad enough so that they may not be suppressed in the SN environment.
While FFCs may occur at various regions across a wide range of densities in the SN environment~\cite{Nagakura:2021hyb, Akaho2024b}, e.g., a thin shell inside the PNS~\cite{Glas2020a}, the decoupling regions where local $\bar\nu_e$ emission dominates due to the multidimensional effect~\cite{Abbar2019b,DelfanAzari:2019epo,Abbar:2020qpi,Nagakura:2019sig}, and the low-density, neutrino free-streaming region due to scatterings~\cite{Morinaga:2019wsv, Abbar:2020qpi}, SFCs may become more relevant in low-density regions.
This is because at such lower densities, FFCs are mostly not expected to be significant and have been suggested to be unlikely to result in significant flavor conversions~\cite{Abbar:2021lmm}. 
Besides FFCs, the collisional flavor instability (CFI)~\cite{Johns2021_1,Padilla-Gay:2022wck,Lin:2022dek,Xiong:2022zqz,Liu:2023pjw} is expected to exist mainly at decoupling densities of $\rho \sim 10^{10}-10^{12} \, {\rm g/cm^3}$~\cite{Johns2021_1,Xiong2023a,Liu2023b}. 
At low densities, even if CFIs exist, they 
are expected to be subdominant compared to SFCs. 
This is because CFIs occur on collisional scales, which are typically $\gtrsim$ a few tens of kilometers in lower-density regions—significantly larger than the SFC characteristic scales of order $\omega^{-1}$. All these indicate that SFC can be the dominant FC of neutrinos in the low-density regions behind the SN shock and can potentially affect SN dynamics, which calls for better understanding of SFC. 
Moreover, SFC are also expected to occur during the PNS cooling phase and can significantly influence SN nucleosynthesis and neutrino signals.

Early pioneering studies of SFC relied on the assumption of a stationary neutrino gas in the SN environment~\cite{Pastor:2002we,Duan:2005cp,Hannestad2006g,duan:2006an,duan:2010bg}. However, to fully capture the potential impact of SFC in low-density regions, it is essential to account for both spatial and temporal evolution of the neutrino flavor field. 
While a few works have performed numerical simulations to study the evolution of different FCs, including SFCs, over static global SN profiles~\cite{Xiong:2022vsy,Shalgar:2024gjt}, they have adopted artificially quenched neutrino-neutrino interaction potentials, which can affect the outcome of true SFC evolution.
In this study, we take a different approach aiming to investigate neutrino flavor equilibration in the context of SFCs for a monochromatic neutrino gas, considering moderate lepton number asymmetries, e.g., $0.7 \lesssim n_{\bar\nu_e}/n_{\nu_e} \lesssim 1.3$ without considering momentum changing collisions. 
To this end, we analyze the evolution of a neutrino gas in periodic boxes and examine the spatially averaged quantities within the box. 
This approach closely follows the methodology used in the study of flavor equilibration in FFCs~\cite{Bhattacharyya2021a,Wu2021a,Richers2021b,Richers2022a,Zaizen2023a,Zaizen2023b,Xiong2023c,Froustey:2023skf,DelfanAzari:2024xgs,Fiorillo:2024qbl,Richers:2024zit,George:2024zxz,Liu:2024nku,Liu:2025tnf}.  
Our motivation for taking this approach for SFCs stems from the fact that, in the regions of interest where $\rho \sim 10^{9}-10^{10} \ {\rm g/cm^3}$, the collisional scales are expected to be significantly larger than the characteristic scales of SFCs, as discussed above. 
Under such conditions, slow modes can effectively be treated as fast compared to the collisional scale, justifying our approach. This consideration is particularly important, as the primary goal of this study is to develop a better understanding of the local quasi-equilibrium state of SFCs. 
We note that although unlike FFCs, the local SFC quasi-equilibrium state cannot be directly used for instantaneous flavor redistribution in global transport or in CCSN simulations, its identification could help develop practical methods to include the effects of FCs in CCSN simulations~\cite{Johns:2024dbe,Nagakura:2023jfi,Xiong:2024pue}.

Our investigations show that the SFCs drive the neutrino gas toward a generic asymptotic state within a few $\omega^{-1}$, in which the subdominant neutrino species reaches an approximate flavor equipartition both in both number density and angular distribution.
Meanwhile, the dominant species approaches a state consistent with lepton number conservation laws. 
In particular, we show that this equilibration is caused by the inhomogeneity-induced decoherence in the neutrino gas. Needless to say, the existence of a generic and simple asymptotic state in the neutrino gas can facilitate the robust incorporation of neutrino FCs into state-of-the-art CCSN simulations.

This paper is structured as follows. In Sec.~\ref{sec:QKE}, we describe the quantum kinetic equations (QKEs) governing the flavor evolution of neutrinos in a dense neutrino gas. Sec.~\ref{sec:simulation} introduces our simulation methods, and in Sec.~\ref{sec:LSA}, we discuss the linear stability analysis of the neutrino gas. We present our main results in Sec.~\ref{sec:res}, before concluding in Sec.~\ref{sec:conc}.

\section{Quantum Kinetic Equations}\label{sec:QKE}

We work under the assumption of two neutrino flavors in a monochromatic neutrino gas, where electron (anti)neutrinos are denoted by $\nu_e (\bar{\nu}_e)$ and the heavy-lepton neutrino flavor admixture by $\nu_x (\bar{\nu}_x)$. 
The flavor content of the system is described by the Wigner-transformed (anti)neutrino density matrices $\rho$ and $\bar{\rho}$ which are complex-valued matrices,

\begin{eqnarray}\label{eq:rho}
    \rho = \begin{pmatrix} 
    \rho_{ee} & \rho_{ex} \\
    \rho_{ex}^{*} & \rho_{xx} 
    \end{pmatrix}, \hspace{3mm}
    \bar{\rho} = \begin{pmatrix} 
    \bar{\rho}_{ee} & \bar{\rho}_{ex} \\
    \bar{\rho}_{ex}^{*} & \bar{\rho}_{xx} 
    \end{pmatrix}
    \ ,
\end{eqnarray}
where the diagonal elements of the density matrices are related to the occupation numbers, while the oﬀ-diagonal ones describe the coherence between flavors.

For simplicity, throughout this work we assume that the neutrino gas possesses axial symmetry around the z-direction. This means that we impose axial symmetry in the initial conditions of the system as well as on the Hamiltonian that describes FC. Unless explicitly stated otherwise, we do not assume spatial homogeneity in either the initial state of the system or its solutions.
We restrict ourselves to the time evolution of 1+1D simulations of FC, which means that we include one spatial coordinate and one angular variable. 
We choose the direction along $\hat{z}$, while the medium remains homogeneous in the $xy$ plane. 
Under these assumptions, the neutrino velocity $\vec{v}$ reduces to its projection along the axis of symmetry, $\vec{v}\cdot \hat{z} = v_z$. Throughout this work, we use natural units ($\hbar=c=1$).

Under the assumptions of axial symmetry and spatial inhomogeneities along the z-direction, the QKEs in the absence of collisions are given by:
\begin{eqnarray}\label{eq:eoms}
    (\partial_t + v_z \partial_z)\rho = -i[H,\rho] \ , \nonumber \\
    (\partial_t + v_z \partial_z)\bar{\rho} = -i[\bar{H},\bar{\rho}] \ , 
\end{eqnarray}
where the Hamiltonian, $H$, receives contributions from the vacuum term, neutrino-matter interactions, and the neutrino-neutrino interactions within the neutrino background, e.g., $H = H_{\rm vac} + H_{\rm mat} + H_{\nu\nu}$. 
The Hamiltonian for antineutrinos is given by $\bar{H} = -H_{\rm vac} + H_{\rm mat} + H_{\nu\nu}$, where the only difference is the $H_{\rm vac} \rightarrow -H_{\rm vac}$ sign change between neutrinos and antineutrinos~\footnote{Note that we have adopted the convention different from that taken in e.g. \cite{duan:2006an,Wu2021a}. 
These two conventions are related by a redefinition of $\bar\rho\rightarrow\bar\rho^*$; see \cite{Johns:2025mlm} for discussions.}. 
In the two flavor approximation, the vacuum Hamiltonian is given by:
\begin{eqnarray}
    H_{\rm vac} = \frac{\omega}{2} \begin{pmatrix} 
    -\cos{2\theta_V} & \sin{2\theta_V} \\
    \sin{2\theta_V} & \cos{2\theta_V} 
    \end{pmatrix} \ ,
\end{eqnarray}
where $\theta_V$ is the vacuum mixing angle and $\omega$ is the vacuum oscillation frequency for which we have $\omega>0$ for normal ordering (NO) and $\omega<0$ for inverted ordering (IO).
The matter contribution to the Hamiltonian can be written as,
\begin{eqnarray}
    H_{\rm mat} = \begin{pmatrix} 
    \lambda & 0 \\
    0 & 0 
    \end{pmatrix} \ ,
\end{eqnarray}
with $\lambda = \sqrt{2}G_{\rm F}\rho_{\rm B} Y_e / m_N$ 
being the matter-neutrino potential, where $G_{\rm F}$ is the Fermi constant, $\rho_{\rm B}$ is the baryon density, $Y_e$ is the electron fraction, and $m_N$ is the nucleon mass. 
Lastly, the neutrino-neutrino Hamiltonian is given by: 
\begin{eqnarray}\label{eq:Hnunu}
    H_{\nu\nu} = \mu \int dv^{\prime}_z [\rho(t,z,v^{\prime}_z)-\bar{\rho}(t,z,v^{\prime}_z)](1- v_z v^{\prime}_z) \ ,
\end{eqnarray}
where $\mu=\sqrt{2}G_F n_{\nu_e}$ describes the strength of the weak interaction potential.
The $H_{\nu\nu}$ term couples neutrinos of different momenta and is responsible for the non-linear nature of neutrino FC. 
Note that we assume both $\mu$ and $\lambda$ are constant in $t$ and $z$ coordinates, hence ignoring any potential effect due to time-variation~\cite{Fiorillo:2024qbl,Liu:2024nku} or spatial inhomogeneity~\cite{Sigl:2021tmj,Bhattacharyya:2025gds} on FCs. 

For practical purposes, here we ignore the constant matter contribution to the Hamiltonian, given the fact that it can be rotated away by frame transformation. 
We then use the matter-suppressed mixing angle~\cite{Esteban-Pretel:2008ovd,Shalgar:2025oht}, which is fixed to be $10^{-5}$ radians. 
We have checked that our results remain qualitatively unchanged once the matter Hamiltonian is included with a representative value of $\lambda=2\mu$ and vacuum mixing angle of $\theta_V = 0.11$ radians. In realistic CCSN environments, $\lambda/\mu$ can be as large as $\sim 10$ behind the shock during the accretion phase~\cite{Chakraborty:2011nf,Bhattacharyya:2025gds} or as small as $\sim 0.1$ between $\sim 50$ and 500~km in the neutrino-driven winds~\cite{Fischer:2023ebq,Bhattacharyya:2025gds}.

Moreover, since $\mu$ and $\omega$ are the only energy scales in the system, we express time and distance in units of $\mu^{-1}$.
Consequently, specifying the ratio $\omega/\mu$ allows us to fully determine the flavor evolution of a particular system.
For the remainder of the paper, we fix $\mu=1$ so that all quantities are shown as dimensionless, with the understanding that their physical units are scaled by the correct power of $\mu$.

\section{Simulation of SFC}\label{sec:simulation}

To study the flavor evolution of the neutrino gas we make use of the publicly available \texttt{COSE}$\nu$ code~\cite{George:2022lwg}, which evolves the components of the neutrino density matrices discretized in space and polar angle assuming axial symmetry. 
\texttt{COSE}$\nu$ offers a finite difference (FD) and a finite volume (FV) method for the treatment of neutrino advection. 
While the FD version of \texttt{COSE}$\nu$ evaluates advection terms using a fourth-order finite difference method with third-order Kreiss-Oliger (KO) artificial dissipation, the FV version of \texttt{COSE}$\nu$ uses a finite difference volume method with seventh-order WENO reconstruction for this purpose. 
Both methods implement a fourth-order Runge-Kutta solver for time integration. 
More details can be found in the code documentation~\cite{George:2022lwg}. 
Moreover, for a robust code comparison among other research groups, we refer the reader to Ref.~\cite{Richers2022a}.

\begin{figure}[tb!]
    \centering
    \includegraphics[width=1.\columnwidth]{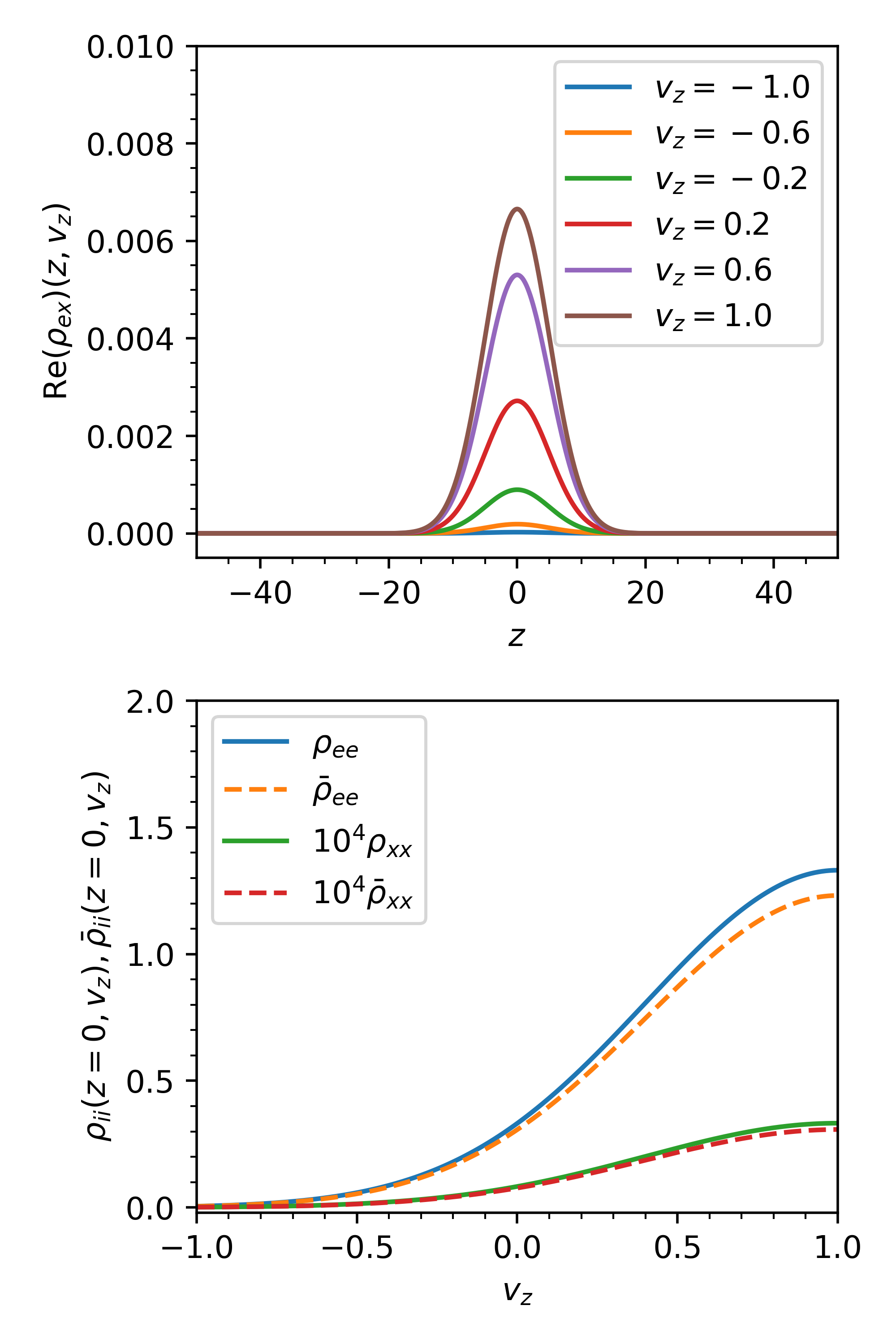}
    \caption{Initial conditions for the matrix elements of $\rho(z,v_z)$ and $\bar{\rho}(z,v_z)$ using $\alpha=0.925$ and $\sigma=0.6$. {\it Upper panel:} Spatial dependence of the real part of the off-diagonal term, ${\rm Re}(\rho_{ex})$, for various angular bins in range $-1 \leq v_z \leq 1$. The spatial distribution is localized around $z=0$ for $-50 \leq z \leq 50$. {\it Lower panel:} Angular distributions of the diagonal terms of the matrix elements at $z=0$. There is no angular crossings between the electron and anti-electron distributions.
    The non-electron flavors $\rho_{xx}$ and $\bar{\rho}_{xx}$ are $\sim 10^{-5}$ at the most forward bin with $v_z=1$.}
    \label{fig:initial}
\end{figure}

Unless otherwise specified, we implement the FD version of \texttt{COSE}$\nu$ to evolve the flavor evolution of our 1+1D simulation inside a box with periodic boundary conditions. We have confirmed that our results do not depend on the chosen size of the box. The benchmark value of the spatial range is set to $z \in [-50,50]$, discretized by $N_z=1000$ equally spaced bins. 
For the angular domain, we take $N_{v_z}=512$ angular bins equally spread in $v_z \in [-1,1]$. With such a large number of $N_{v_z}$, the system is free from late-time numerical artifacts that arise when too few angular bins are used to represent the continuous angular distribution of (anti)neutrinos inaccurately. Moreover, the spatial resolution is high enough to resolve relevant wavelengths of the system (larger than $L/N_z = 0.1$). The Courant–Friedrichs–Lewy factor and the KO dissipation parameter are chosen to be $0.4$ and $0.1$, respectively. 

The initial conditions of $\rho$ and $\bar\rho$ depend on the velocity $v_z$ and the spatial coordinate $z$ and are set as described in the following. 
We assume the neutrino gas consists of monochromatic $\nu_e$ and $\bar\nu_e$ before applying a spatially dependent perturbation (see below). 
The normalized angular distribution of neutrinos is taken to be uniform in $z$ and is parametrized by:
\begin{eqnarray}
    g(v_z, \sigma) = A e^{-(v_z-1)^2/2\sigma^2} \ ,
\end{eqnarray}
where $\sigma$ is the width of a half-Gaussian distribution and $A^{-1} = \sigma \sqrt{\pi/2} \ {\rm erf} (\sqrt{2}/\sigma)$ with ${\rm erf}$ the error function. 
For antineutrinos, we assume their angular distribution follows $\bar g (v_z,\sigma) = \alpha g(v_z,\sigma)$ with $\alpha = n_{\bar{\nu}_e}/n_{\nu_e}$ being the neutrino-antineutrino asymmetry factor.
The choice of the same width parameter $\sigma$ for neutrinos and antineutrinos is to avoid appearance of ELN crossings that can trigger FFCs. In the absence of an ELN angular crossing, no unstable modes exist in the limit $\omega \rightarrow 0$, and any instability is due to SFC. In the case of SFC, a crossing in the energy spectrum is required for modes to become unstable. In a single-energy setup, this is typically achieved by treating antineutrinos as having “negative” energy ($\omega<0$), since they pick up a minus sign in $H_{\rm vac}$ compared to neutrinos ($\omega>0$), thus guaranteeing a crossing in the energy spectra.

\begin{figure}[tb!]
    \centering
    \includegraphics[width=1.\columnwidth]{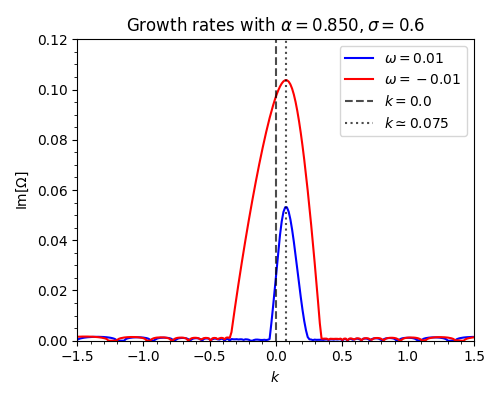}
    \caption{Imaginary part of the eigenvalue $\Omega$ as a function of the wave-number, $k$, for both mass orderings. 
    The angular distribution parameters are fixed to $\alpha=0.850$ and $\sigma=0.6$ (see Fig.~\ref{fig:initial}).
    The IO case (red) has larger growth rates and thus develops faster than the NO case (blue), which is visible in the $t\simeq 100$ snapshot showed in Fig~\ref{fig:P3_on_v_z_plane}. 
    Moreover, one can see that the $k=0$ mode (dashed) is subdominant in the NO case, which means that it is mainly driven by $k\neq0$ modes, such as the dominant mode $k=0.075$ (dotted). 
    On the other hand, in the IO case the $k=0$ mode is approximately the dominant one.}
    \label{fig:lsa_NO_IO}
\end{figure}

\begin{figure}[tb!]
    \centering
    \includegraphics[width=1.\columnwidth]{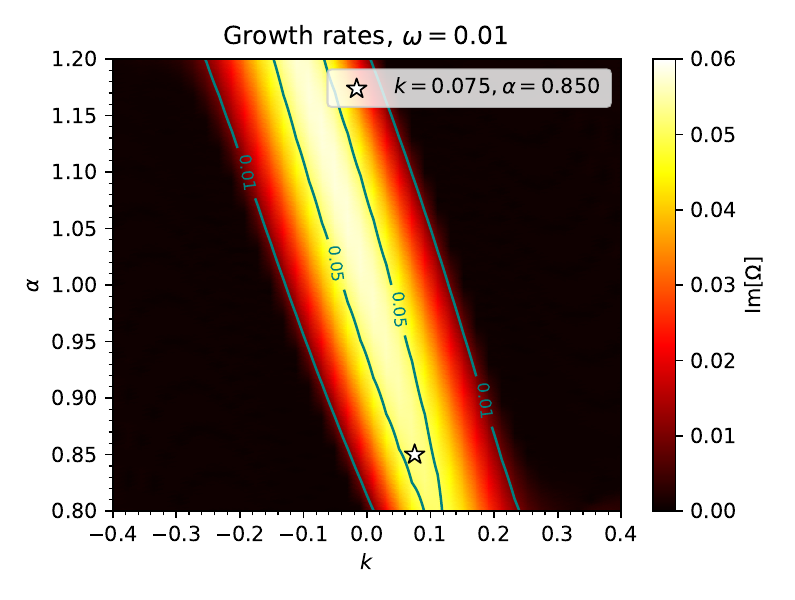}
    \includegraphics[width=1.\columnwidth]{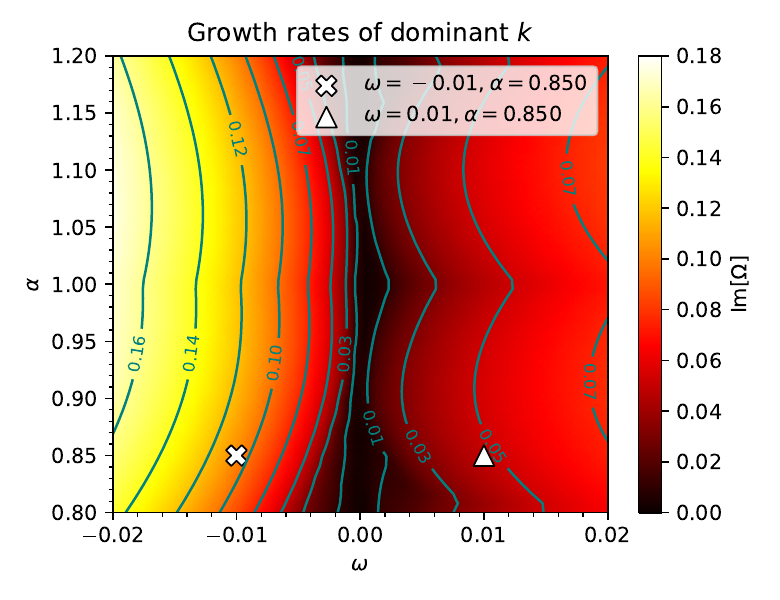}
    \caption{{\it Top:} Growth rate $\mathrm{Im}(\Omega)$ as a function of the wave-number $k$ and the asymmetry parameter $\alpha$ while keeping the vacuum frequency fixed to $\omega=0.01$ (NO). 
    The star marker corresponds to the largest growth rate marked by the vertical dotted line in Fig~\ref{fig:lsa_NO_IO}. 
    {\it Bottom:} Growth rate $\mathrm{Im}(\Omega)$ of the dominant $k$ mode as a function of the vacuum frequency $\omega$ and the asymmetry parameter $\alpha$. 
    The triangle and cross markers show the growth rates for the NO and IO systems shown in Fig~\ref{fig:lsa_NO_IO}.
    }
    \label{fig:lsa_colormaps}
\end{figure}

In principle, one could simply take spatially homogeneous $\rho_{ee}=g(v_z,\sigma)$ and $\bar\rho_{ee}=\bar g(v_z,\sigma)$ to be the only non-zero component of density matrices as the initial condition for the system to evolve, since the vacuum term would be seeding the off-diagonal elements to trigger SFCs. 
However, as the vacuum term can only trigger the unstable homogeneous mode, and physical effects originating from the inhomogeneous modes would not emerge until much later times through possible tiny numerical rounding errors.
Thus, we apply a spatially dependent perturbation seed to the initial real parts of the density matrices by  
${\rm Re}(\rho_{ex}) = g(v_z,\sigma) \varepsilon/2$, ${\rm Re}(\bar{\rho}_{ex}) = \bar{g}(v_z,\sigma) \varepsilon/2$, where  
\begin{eqnarray}
    \varepsilon (z,z_0) = B e^{-(z-z_0)^2/50} \ .
\end{eqnarray}
Unless otherwise specified, for numerical examples we fix $B = 0.01$, $z_0=0$, and choose the denominator of the exponential to be 50 which corresponds to a Gaussian width of 5. 
For the complex parts, ${\rm Im}(\rho_{ex}) = {\rm Im}(\bar{\rho}_{ex}) = 0$ initially. 
Given these, the initial diagonal elements of $\rho$ and $\bar\rho$ are given by  
\begin{eqnarray}\label{eq:angular_nu}
    \rho_{ee} &=&  g(v_z,\sigma) (1+\sqrt{1-\varepsilon^2})/2 \ ,  \\ 
    \rho_{xx} &=& g(v_z,\sigma) (1-\sqrt{1-\varepsilon^2})/2 \ ,
\end{eqnarray}
and 
\begin{eqnarray}\label{eq:angular_anu}
    \bar{\rho}_{ee} &=& \bar{g}(v_z,\sigma) (1+\sqrt{1-\varepsilon^2})/2 \ , \\ 
    \bar{\rho}_{xx} &=& \bar{g}(v_z,\sigma) (1-\sqrt{1-\varepsilon^2})/2 \ , 
\end{eqnarray}
which introduces a small amount of heavy-lepton neutrinos and antineutrinos in the initial condition.
We show in Fig.~\ref{fig:initial} the initial conditions for the (anti)neutrino matrix elements following the aforementioned parametrization with $\alpha=0.925$.

Additionally, we can use the Pauli matrices to map the neutrino and antineutrino density matrices onto vectors in flavor space, whose components $P_i$ are:
\begin{eqnarray}
    P_1 &\coloneqq& {\rm Tr}(\rho \sigma^x)/g(v_z,\sigma)  = 2 {\rm Re}(\rho_{ex})/g(v_z,\sigma)  \ , \nonumber \\
    P_2 &\coloneqq& {\rm Tr}(\rho \sigma^y)/g(v_z,\sigma) = -2 {\rm Im}(\rho_{ex})/g(v_z,\sigma)  \ , \nonumber \\
    P_3 &\coloneqq& {\rm Tr}(\rho \sigma^z)/g(v_z,\sigma)  = (\rho_{ee}-\rho_{xx})/g(v_z,\sigma)  \ ,
\end{eqnarray}
where a similar definition follows for antineutrinos with the replacement $g(v_z,\sigma)\rightarrow \bar{g}(v_z,\sigma)$. 
Due to unitarity, the norm of the total polarization vector $\vec{P}=(P_1,P_2,P_3)$ has to remain constant in time along the trajectory of neutrinos. 
This notation, in particular, allows for simple definition of some useful quantities such as the electron-minus-heavy-lepton neutrino number [(E-X)LN], $P_3-\bar{P_3}$, and the complex transverse quantity $s = P_1 - i P_2$ which captures the onset of unstable solutions in flavor space.

\begin{figure*}[tbh!]
    \centering
    \includegraphics[width=.97\textwidth]{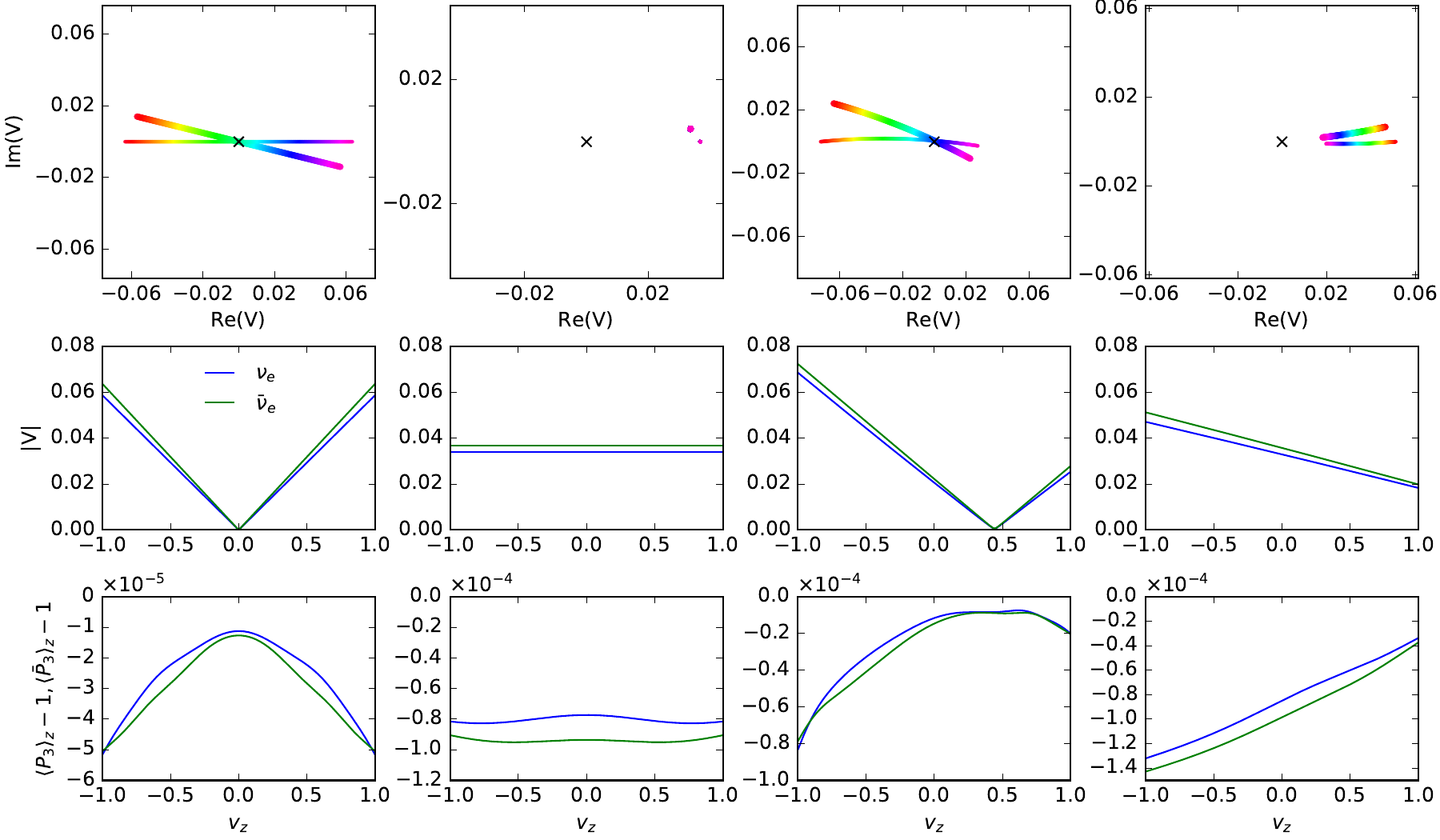}
    \llap{\parbox[b]{11.5in}{\small isotropic, NO\\\rule{0ex}{3.7in}}}
    \llap{\parbox[b]{8.3in}{\small isotropic, IO\\\rule{0ex}{3.7in}}}
    \llap{\parbox[b]{5.0in}{\small anisotropic, NO\\\rule{0ex}{3.7in}}}
    \llap{\parbox[b]{1.8in}{\small anisotropic, IO\\\rule{0ex}{3.7in}}}
    \caption{Distributions (upper), absolute values of the eigenvectors $\vec V$ (middle), and a simulation snapshot of $\langle P_3\rangle$ and $ \langle\bar{P}_3\rangle$ in linear regime (bottom). 
    The columns correspond to isotropic angular distributions ($\sigma\rightarrow \infty$, $k=0$) for NO and IO, and anisotropic angular distributions ($\sigma=0.6$, $k=0.075$) for NO and IO, respectively. 
    We keep the parameters $\alpha=0.85$, and $|\omega|=0.01$ fixed. 
    The normalization is such that $\sum |{\vec V}|^2 = 1$. 
    In the upper panels, the color distinguishes the angular bins from $-1$ (red) to $1$ (purple), while thick and thin markers correspond to $\nu_e$ and $\bar\nu_e$, respectively.
    The bottom panels show the angular distributions of the averaged $P_3$ and $\bar{P}_3$ in the linear regime ($t=60$ for NO, and $t=20$ for IO). 
    The angular modes at which $P_3,\bar{P}_3$ are closest to 1 correspond to the locations where $|{\vec V}|$ has a minimum. 
    The shape of the angular distribution in the linear regime is described by the shape of $|{\vec V}|$.
    }
    \label{fig:eigen_vectors}
\end{figure*}

\section{Linear Stability Analysis}\label{sec:LSA}

We perform the linear stability analysis~\cite{Izaguirre2017a} on the system introduced in Section~\ref{sec:QKE} to identify the regions of the parameter space that are prone to flavor instabilities.

The density matrices in Eq.~\ref{eq:rho} can be written as
\begin{eqnarray}
    \rho &=& \begin{pmatrix} 
    \rho_{ee}(t_0)-\Delta(t) & \rho_{ex}(t) \\
    \rho_{ex}^{*}(t) & \rho_{xx}(t_0)+\Delta(t) 
    \end{pmatrix} \ , \\
    \bar{\rho} &=& \begin{pmatrix} 
    \bar{\rho}_{ee}(t_0)-\bar{\Delta}(t) & \bar{\rho}_{ex}(t) \\
    \bar{\rho}_{ex}^{*}(t) & \bar{\rho}_{xx}(t_0)+\bar{\Delta}(t) 
    \end{pmatrix} \ ,
\end{eqnarray}
where we have omitted the explicit dependence of the matrix elements on $v_z$ and $z$ for readability. Here, the components $\rho_{ex}$ and $\bar{\rho}_{ex}$ are initially small compared to the diagonal terms. 
Thus, the equations of motion (Eqs.~\ref{eq:eoms}) can be expanded in series in $\rho_{ex}$ and $\bar{\rho}_{ex}$, and one can immediately realize that the equation for $\Delta$ goes like $\mathcal{O}(\rho_{ex}^2)$. 
This implies that the evolution of the diagonal terms remains subleading until the off-diagonal terms have fully developed. In flavor space, to linear order, the motion of the polarization vector is restricted to the plane whose normal direction points to the flavor $z$-direction. 
By focusing on the leading order, we neglect the evolution of $\Delta$, and concentrate on the equations of motion for the off-diagonal terms:
\begin{eqnarray}\label{eq:linearEoM}
    i (\partial_t + v_z\partial_z) \rho_{ex} &=& -\omega\rho_{ex} + H_{ee}\rho_{ex} - \rho_{ee}H_{ex} \ , \\ 
    i (\partial_t + v_z\partial_z) \bar{\rho}_{ex} &=& +\omega\bar{\rho}_{ex} + H_{ee}\bar{\rho}_{ex} - \bar{\rho}_{ee}H_{ex} \ ,
\end{eqnarray}
where we have once again assumed that the non-electron flavors initially vanish. 
In Eqs.~\eqref{eq:linearEoM} and what follows in this subsections, $H_{\alpha\beta}$ refers to the $\alpha\beta$ component of $H_{\nu\nu}$.  
Moreover, we remind the reader that $H_{\nu\nu}=\bar{H}_{\nu\nu}$. 
Under the plane-wave ansatz of the normal mode solution for $\rho_{ex}, \bar{\rho}_{ex} \sim  e^{i k v_z}$
the equation reads:
\begin{eqnarray}
    i \partial_t \rho_{ex} &=& -\omega\rho_{ex} + v_z k \rho_{ex} + H_{ee}\rho_{ex} - \rho_{ee}H_{ex} \ , \\ 
    i \partial_t \bar{\rho}_{ex} &=& +\omega\rho_{ex} + v_z k \bar{\rho}_{ex} + H_{ee}\bar{\rho}_{ex} - \bar{\rho}_{ee}H_{ex} \ .
\end{eqnarray}

The equation of motion for the off-diagonal terms of neutrinos and antineutrinos can be written all together in matrix form:
\begin{eqnarray}\label{eq:partial_t_V}
    i \partial_t \vec{V} = \Lambda \vec{V} \ ,
\end{eqnarray}
where $\Lambda$ is a $2N_{v_z} \times 2N_{v_z}$ matrix and the vector $\vec{V}$ contains the off-diagonal terms for all the velocity modes $v_i = v_0, ..., v_{N_{v_z}-1}$ for neutrinos and antineutrinos,
\begin{eqnarray}
    \vec{V} = \begin{pmatrix} 
    \rho_{ex,0} & \bar{\rho}_{ex,0} & ... & \rho_{ex, N_{v_z}-1} & \bar{\rho}_{ex, N_{v_z}-1}
    \end{pmatrix}^{T} \ ,
\end{eqnarray}
where the short notation for $\rho_{ex,0} \equiv \rho_{ex}(v_0)$ refers to the off-diagonal component of the neutrino density matrices for the mode $v_0$. It is clear that Eq.~\ref{eq:partial_t_V} is now an eigenvalue problem.

The elements of $H_{ee}$ are computed from Eq.~\ref{eq:Hnunu} in a discretized fashion as,
\begin{equation}
    H_{ee, i} = \Delta v \sum_{j} \left( \rho_{ee, j} - \bar{\rho}_{ee, j} \right) (1 - v_i v_j) \ ,
\end{equation}
where $v_i$ represents the velocity component, $\rho_{ee, j}$ and $\bar{\rho}_{ee, j}$ are input distributions, and $\Delta v$ is the velocity discretization step. The indices take the values $i,j =  0,1,...,N_z-1$.
The components of the matrix $\Lambda$ can be constructed by computing,

\begin{eqnarray}\label{eq:Lambda}
    \Lambda_{2i, 2j} &=& -\rho_{ee, i} \Delta v (1 - v_j v_i) + \delta_{ij} (H_{ee, i} - \omega + v_i k) \ ,  \nonumber \\
    \Lambda_{2i+1, 2j+1} &=& +\bar{\rho}_{ee, i} \Delta v (1 - v_j v_i) + \delta_{ij}(H_{ee, i} + \omega + v_i k) \ , \nonumber \\
    \Lambda_{2i, 2j+1} &=& +\rho_{ee, i} \Delta v (1 - v_j v_i) \ , \nonumber \\
    \Lambda_{2i+1, 2j} &=& -\bar{\rho}_{ee, i} \Delta v (1 - v_j v_i) \ , 
\end{eqnarray}
where $\delta_{ij}$ is the Kronecker delta which accounts for the extra contribution to the diagonal terms. The matrix $\Lambda$ is completely determined by the initial conditions of the system and does not depend on the dynamical variable $\rho_{ex}$. We compute the eigenvalues of $\Lambda$ and search for the eigenvalue with the largest positive imaginary component. 
The imaginary part of this $\Omega$ represents the growth rate of the unstable solution, or in other words, $\vec{V} \sim e^{{\rm Im}{(\Omega)}t}$ grows at a rate fixed by Im$(\Omega)$.

In Fig.~\ref{fig:lsa_NO_IO}, we show the dispersion relation of a system with $\omega=\pm 0.01$ and (anti)neutrino angular distributions defined by Eqs.~\ref{eq:angular_nu} and~\ref{eq:angular_anu} with $\alpha=0.850$ and $\sigma=0.6$. 
The NO case (blue) generically shows smaller growth rates than the IO case (red) across the $k$-domain. 
Moreover, the IO case has a wider range of instability in $k$ than the NO case. Both mass orderings have a global maximum around the $k=0.075$ mode which is marked by the dotted line. 
Additionally, one can see that the IO case is approximately dominated by the $k=0$ mode, while the NO case has a subdominant homogeneous mode and is thus driven by $k\neq 0$ modes. 
To paint a more general picture of the instability region, in Fig.~\ref{fig:lsa_colormaps} we show the growth rates of the parameter space comprised by $0.8 \leq \alpha \leq 1.2$ and $-0.02 \leq \omega \leq 0.02$. 
In the top panel of Fig.~\ref{fig:lsa_colormaps}, one can see that for the NO case, the systems are more strongly dominated by inhomogeneous ($k \neq 0$) modes when $\alpha$ deviates from 1 as previously discussed.
Moreover, the growth rates are approximately symmetric around $\alpha=1$, e.g., the growth rates of mode $k$ for $\alpha=0.9$ is roughly the same as that of mode $-k$ for $\alpha=1.1$. 
The bottom panel of Fig.~\ref{fig:lsa_colormaps} shows the instability region for the dominant $k$ mode for any given pair of parameters $(\omega,\alpha)$. 
One can see that the $\omega<0$ (IO) quadrant has generally larger growth rates that span a wider range of values of $\alpha$ than $\omega>0$ (NO).

\begin{figure*}[t!]
    \centering
    \includegraphics[width=1.\textwidth]{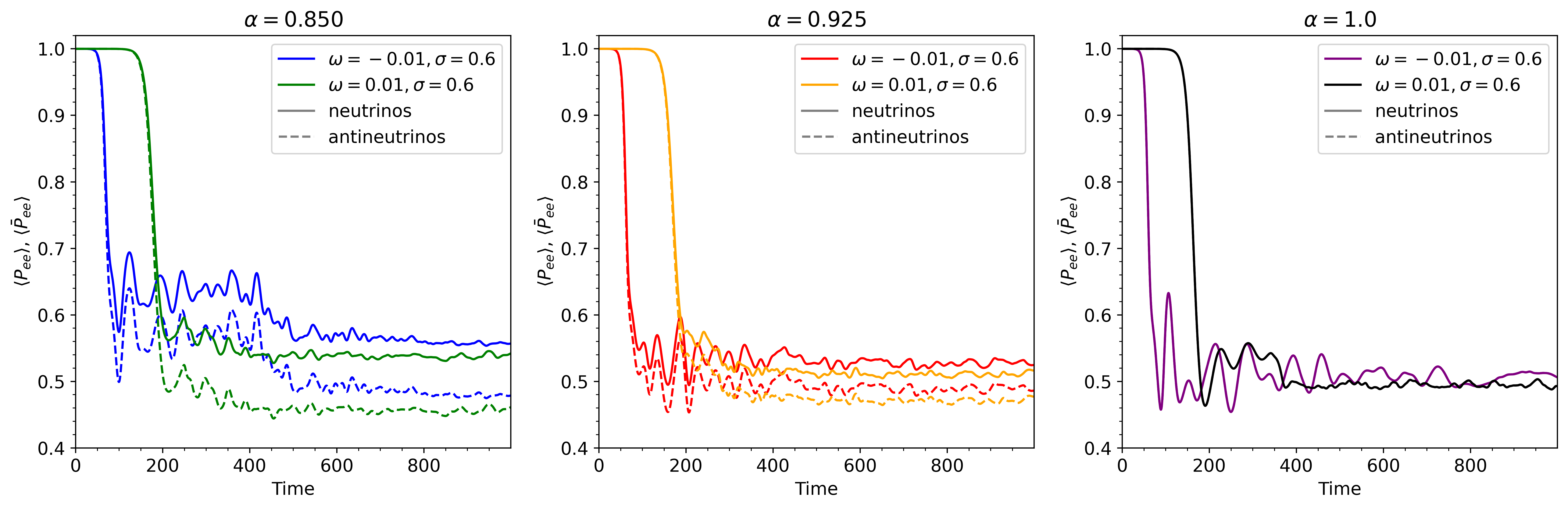}
    \caption{Neutrino and antineutrino survival probabilities averaged over space and angle, here denoted by $\langle P_{ee} \rangle$ (solid line) and $\langle \bar{P}_{ee} \rangle$ (dashed line), as a function of time. 
    The panels show the flavor evolution in the periodic box for $\alpha=\{0.850,0.925,1.0\}$ and $\omega=\{-0.01,0.01\}$, while holding the other parameters constant, such as $\sigma=0.6$. The onset of SFC occurs earlier for $\omega < 0$ (IO), in agreement with our estimates of the instability growth rate (see Sec.~\ref{sec:LSA}). 
    The empirical estimate of the flavor state once the quasi-steady-state is reached is shown in Fig.~\ref{fig:alpha_plot}.
    Despite initial differences, the inhomogeneity-induced decoherence effect leads to very similar quasi-stationary states in both mass orderings. All systems considered have reached flavor equilibration before $t \approx 1000 = 10\ \omega^{-1}$.
    }
    \label{fig:three_alphas}
\end{figure*}

The linear stability analysis not only yields the growth rate of the stability but also provides insights on the late-time evolution by examining specific angular distributions of the unstable modes \cite{PadillaGay2022a,Xiong2023b}.
The panels in the upper two rows of Fig.~\ref{fig:eigen_vectors} compares the major properties of the unstable mode as functions of $v_z$ for $\sigma=0.6$ and $k=0.075$ to the corresponding isotropic cases with infinitely large $\sigma$ so that the unstable mode with maximum growth rate occurs at $k=0$.
With the isotropic angular distribution, the IO unstable mode is independent of $v_z$, which is historically called the bipolar mode~\cite{Duan:2005cp,Hannestad2006g}. 
For NO, the unstable mode exhibits strong angular dependence, referred to as the multi-zenith-angle (MZA) mode~\cite{Raffelt:2007yz}. 
The distribution of the MZA unstable eigenvector in the complex plane forms two lines, each of which corresponds to the components associated with $\rho_{ex}$ and $\bar\rho_{ex}$, since $\vec  V$ is a linear function of $v_z$. These two lines cross the origin at mode $v_z=0$. Moreover, the emergence of multi-azimuthal-angle (MAA) modes is also possible when azimuthal symmetry is broken~\cite{Raffelt2013a}; however, we defer the exploration of MAA to future work, as this study focuses on 1+1D simulations.

The eigenvectors for the anisotropic distribution with $\sigma = 0.6$ differ in shape from those in the isotropic case. For example, the velocity mode with $|\vec V|=0$ is shifted to $v_z \sim 0.4$ in NO case. The IO unstable mode also exhibits $v_z$ dependence. 
Nonetheless, some features remain qualitatively similar to the isotropic cases. For instance, the NO eigenvector still goes through the origin while the IO eigenvector does not.

The above features of the unstable eigenvectors are consistent with the evolution of the spatially-averaged projected polarization of neutrinos $\langle P_3 \rangle_z$ and antineutrinos $\langle \bar P_3 \rangle_z$ in the linear regime when dynamically solving Eq.~\ref{eq:eoms} as shown in the bottom row of Fig.~\ref{fig:eigen_vectors}.
For both NO cases, less flavor conversions occur at the corresponding velocity mode where the eigenvector goes across the origin in the complex plane of linear stability analysis, compared to other velocity modes.
The maximum values of $\langle P_3 \rangle_z$ are at $v_z=0$ and $v_z\sim 0.3$--0.6 for isotropic and anisotropic cases, respectively.
In contrast, less angular dependence of flavor conversion is observed for the isotropic IO case.
Although the distribution of $\langle P_3 \rangle_z$ in the anisotropic IO case exhibits angular dependence, the distribution is monotonic and peaks at $v_z=1$, unlike the $\Lambda$-like shapes in the NO cases where the maximum is at intermediate velocity mode.

\section{Flavor Equilibration of SN Neutrinos Induced by Slow Modes}\label{sec:res}

\begin{figure}[tb!]
    \centering
    \includegraphics[width=0.95\columnwidth]{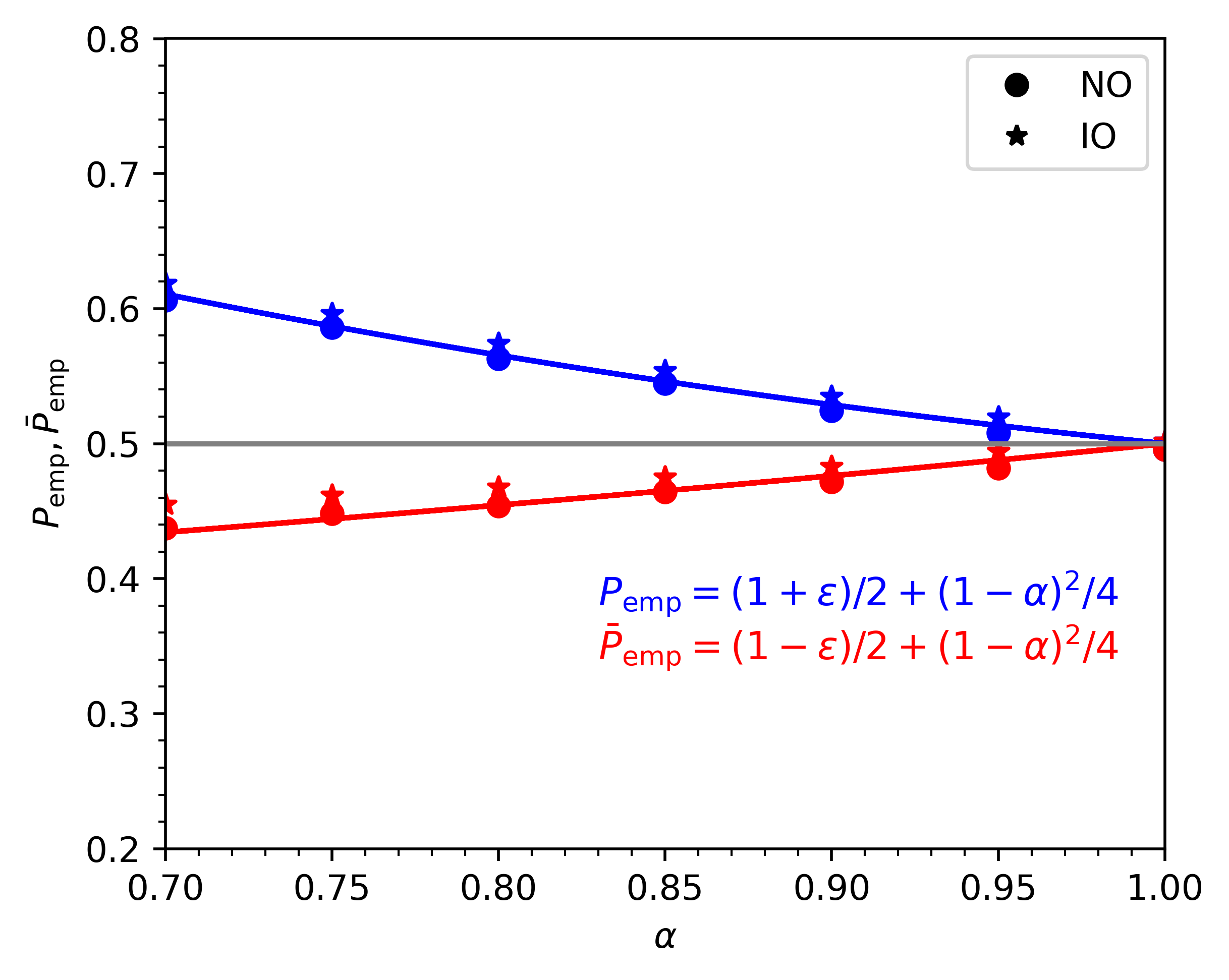}
    \caption{Empirical steady-state of $\langle P_{ee} \rangle$ and $\langle \bar{P}_{ee} \rangle$ as a function of the asymmetry parameter $\alpha$. 
    Blue (red) markers show the final state of neutrinos (antineutrinos) from simulation data. The circle (star) markers correspond to the NO (IO) scenario. 
    We note that higher order corrections that depend on $\omega$ are not included in the present parametrization of $P_{\rm emp}, \bar{P}_{\rm emp}$, but could contribute as much as a few percent to the final steady-state of the system, specially for small $\alpha$.
    } 
    \label{fig:alpha_plot}
\end{figure}

\begin{figure}[tbh!]
    \centering
    \includegraphics[width=0.95\columnwidth]{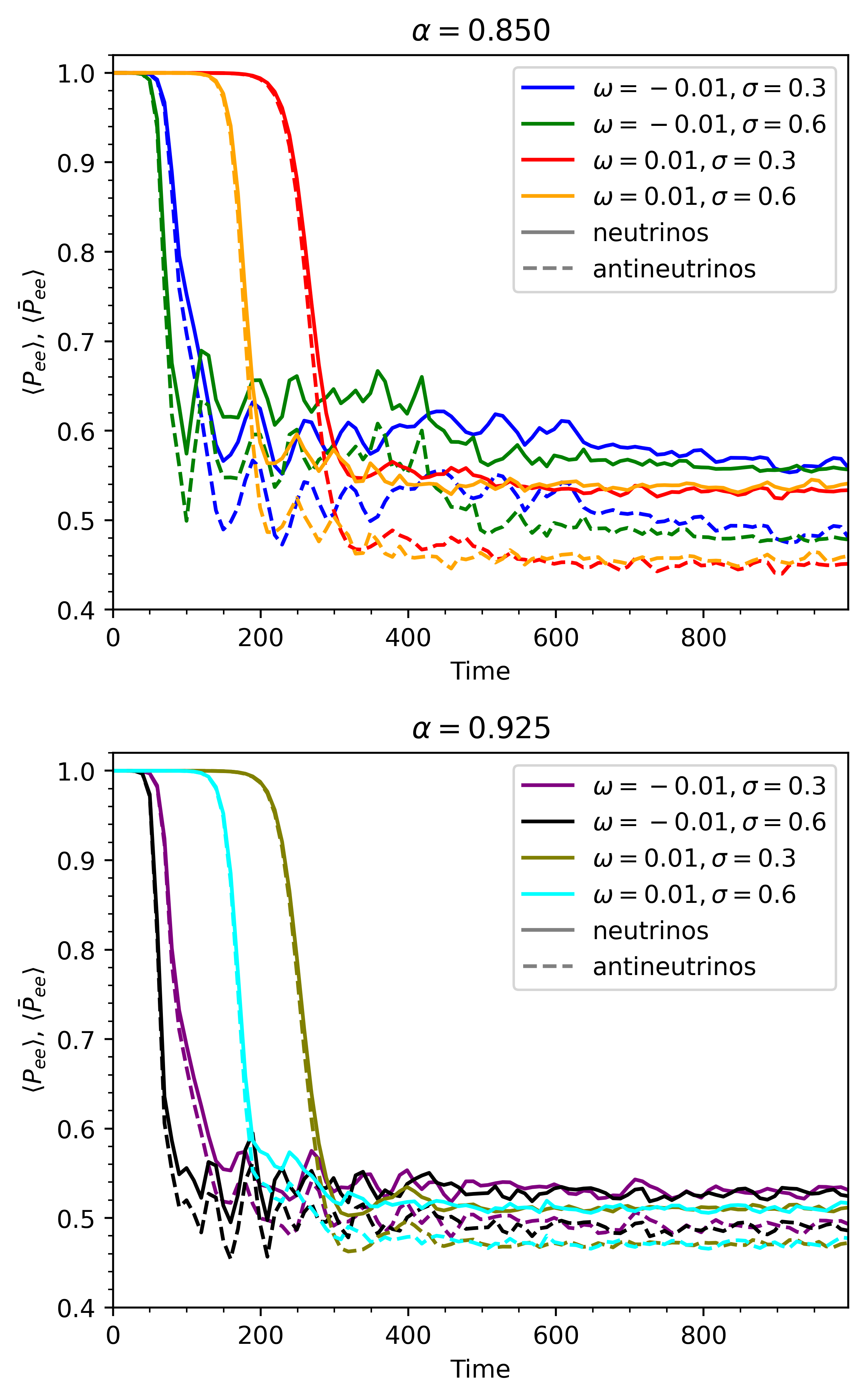}
    \caption{Same as Fig.~\ref{fig:three_alphas}, but additionally showing results from varying the (anti)neutrino angular distribution width, $\sigma$. 
    The flavor evolution and the final flavor state of the neutrino gas in the quasistationary state does not depend on how non-isotropic the distributions are. 
    This is reflected by the fact that the more non-isotropic case ($\sigma=0.3$) although showing strong differences initially ($t<600$), catches up to the same level of conversion at $t=1000$. 
    This also reflects the fact that our empirical description of the final state should not depend on the initial angular distribution of neutrinos. 
    }
    \label{fig:two_sigmas}
\end{figure}

In this section, we illustrate the occurrence of neutrino flavor equilibration in a dense neutrino gas induced by SFCs, for $0.7 \lesssim \alpha \lesssim 1.3$. We begin by presenting results for the case $\alpha \leq 1$ and develop an intuitive understanding of the underlying dynamics. 
Subsequently, we demonstrate that flavor equilibration also occurs for $\alpha>1$, highlighting the generality of this phenomenon.

\smallskip
\textbf{\textit{Flavor equilibration for $\boldsymbol{\alpha \leq 1}$.---}}As illustrated in Fig.~\ref{fig:three_alphas}, the averaged survival probabilities for neutrinos and antineutrinos reach equilibration after a few $\omega^{-1}$ in both neutrino mass orderings. 
While the antineutrinos approach a state closer to equipartition, the neutrinos settle at a value consistent with the conservation of the total (E--X)LN (which is guaranteed to a good degree given the suppressed effective vacuum mixing angle; see Appendix~\ref{appendix:conservedQ}). 
Due to the approximate conservation of (E--X)LN, the final flavor conversion of the neutrinos is related to that of the antineutrinos through
\begin{eqnarray}\label{eq:avePee}
    \langle P_{ee} \rangle \simeq 1-\alpha(1-\langle \bar{P}_{ee} \rangle)  \ ,
\end{eqnarray}
which we have checked to hold in our simulations (see Fig.~\ref{fig:alpha_plot}). 
The occurrence of such a generic equilibration is the main finding of this study. 
This conclusion is further supported by the results presented in Fig.~\ref{fig:two_sigmas}, which demonstrate that both the occurrence of the equilibration and the asymptotic state are independent of the shape of the angular distribution. 
Although different neutrino angular distributions lead to distinct initial dynamics in the neutrino gas, the final asymptotic state remains unaffected by the specific form of the distribution and initial conditions.

In Fig.~\ref{fig:alpha_plot}, we present our empirical predictions for the steady-state survival probabilities
\begin{eqnarray}\label{eq:emp}
    P_{\rm{emp}}(\alpha) &=& \frac{1+\epsilon}{2} + \frac{(1-\alpha)^2}{4} \ , \nonumber  \\ 
    \bar{P}_{\rm{emp}}(\alpha) &=& \frac{1-\epsilon}{2} + \frac{(1-\alpha)^2}{4} \ , 
\end{eqnarray}
which assume the absence of $\nu_x$ flavors in the initial state of the system. 
The function $P_{\rm{emp}}$ ($\bar{P}_{\rm{emp}}$) denotes the empirical steady-state of the $\nu_e$ ($\bar{\nu}_e$) survival probability and is shown with a solid blue (red) line. 
In the estimation of empirical steady state, we have defined the quantity $\epsilon \equiv |1-\alpha|/(1+\alpha)$, which enters in the first term of $P_{\rm{emp}}$ and $\bar{P}_{\rm{emp}}$ with a plus (minus) sign for neutrinos (antineutrinos), while the quadratic term in $\alpha$ is the same for both. 
The markers in Fig.~\ref{fig:alpha_plot} represent the steady-state values of the survival probabilities for systems with an asymmetry parameter between $0.7\lesssim \alpha \lesssim 1.0$. In particular, we find excellent agreement between our numerical simulations and the predicted steady-state values obtained with the empirical functions of Eq.~\ref{eq:emp}.
Moreover, we have verified that the numerical results with an asymmetry parameter between $1.0\lesssim \alpha\lesssim 1.3$ are also in good agreement with the empirical functions. Thus, in the range $0.7 \lesssim \alpha \lesssim 1.3$, our empirical formula provides an accurate approximation of the coarse-grained survival probabilities. While the empirical values for antineutrinos exhibit slight deviations from 0.5, $\bar{P}_{\rm{emp}}$ remains close to 0.5 over the interval $0.7 \lesssim \alpha \lesssim 1.0$.

\begin{figure}[tb!]
    \centering
    \includegraphics[width=0.95\columnwidth]{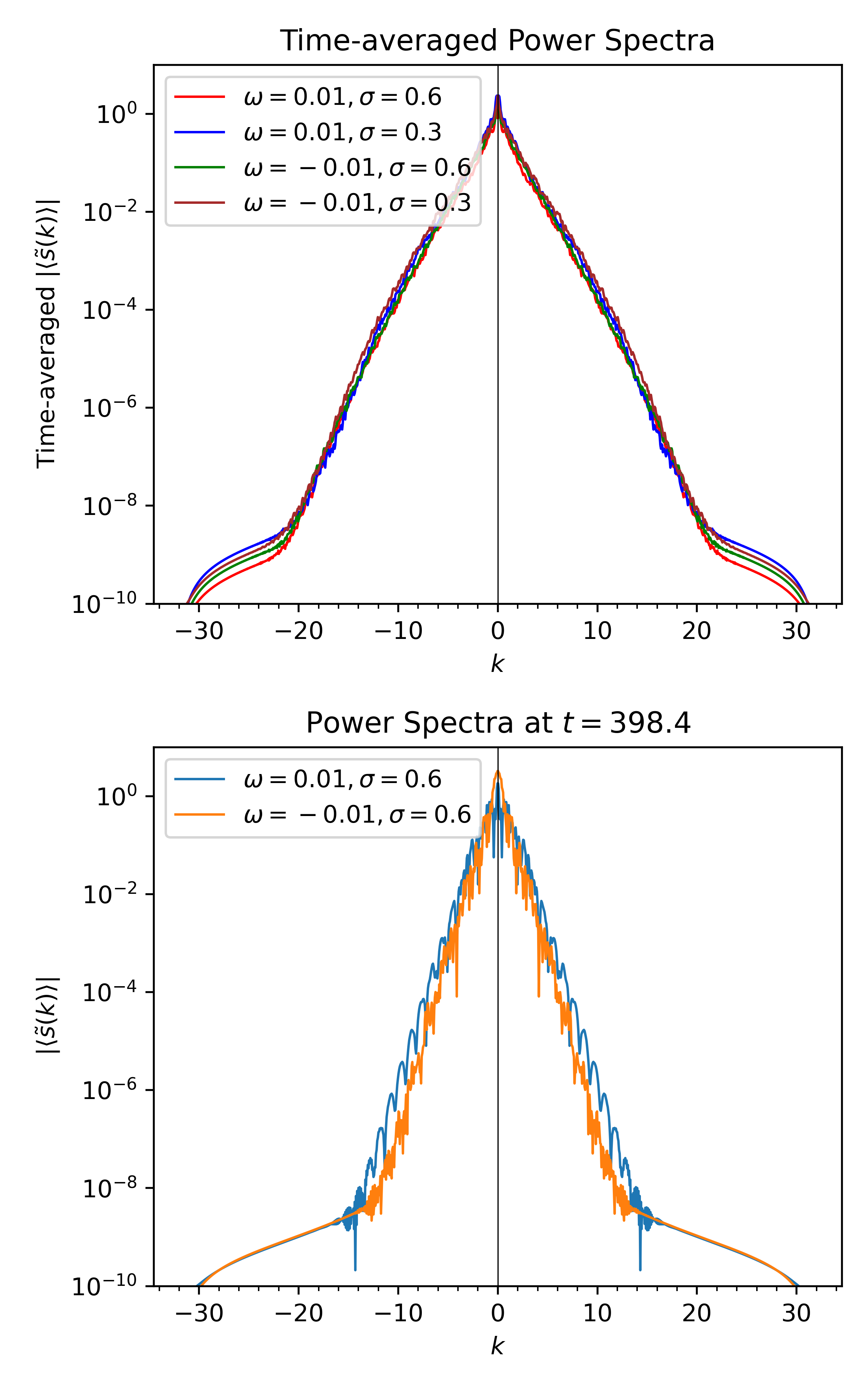}
    \caption{Fourier power spectrum of the off-diagonal component, $s = P_1 - i P_2$. 
    We show the magnitude of the transverse vector in Fourier space, which is a function of $k$, the norm of the wave vector. {\it Top:} Time-averaged power spectrum of the cases with $\sigma=0.3, 0.6$ and $\alpha=0.925$, in both mass orderings. 
    All simulations qualitatively show the same power behavior, where the peak and the slope of the exponential tails are the same regardless of the initial angular conditions and the mass ordering. 
    The time-average is taken in the interval $t = [700,1000]$, within which the unstable modes have fully developed and saturated. 
    {\it Bottom:} Snapshot of the Fourier power spectrum of the $\sigma=0.6$ cases at $t=398.4$.
    }
    \label{fig:power_spectrum}
\end{figure}

\begin{figure*}[ht]
    \centering
    \includegraphics[width=0.49\textwidth]{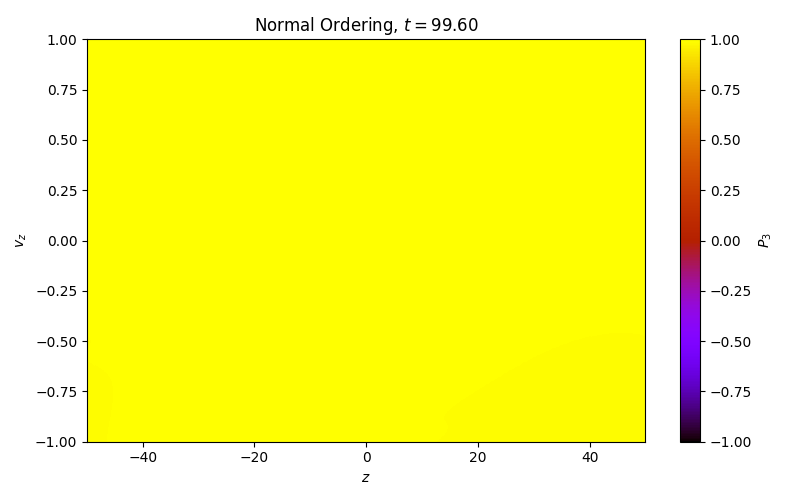}
    \includegraphics[width=0.49\textwidth]{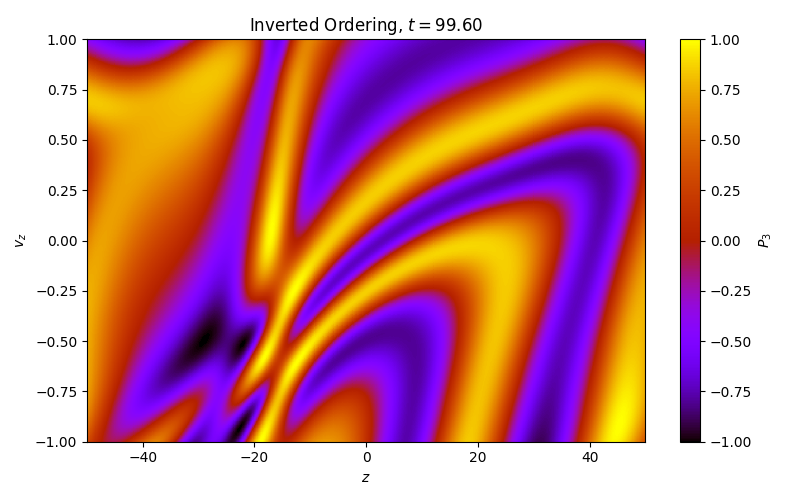}
    \includegraphics[width=0.49\textwidth]{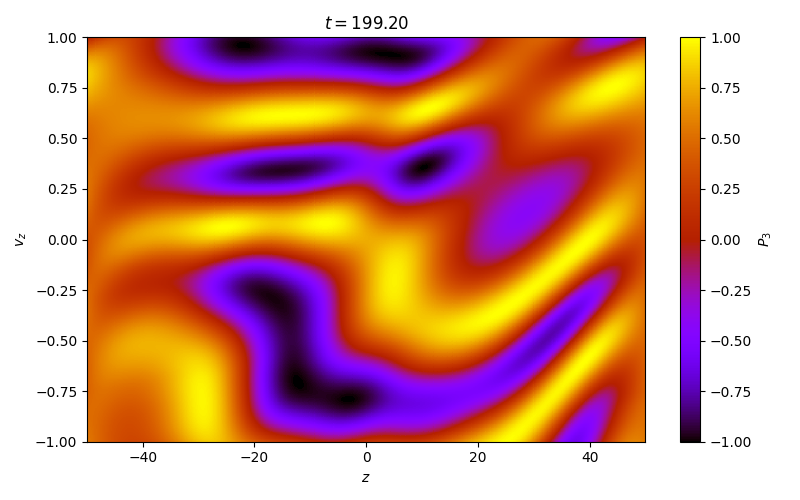}
    \includegraphics[width=0.49\textwidth]{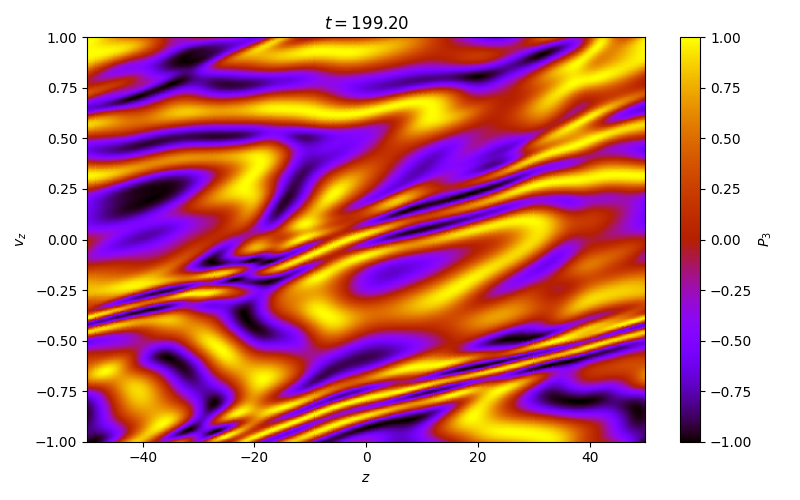}
    \includegraphics[width=0.49\textwidth]{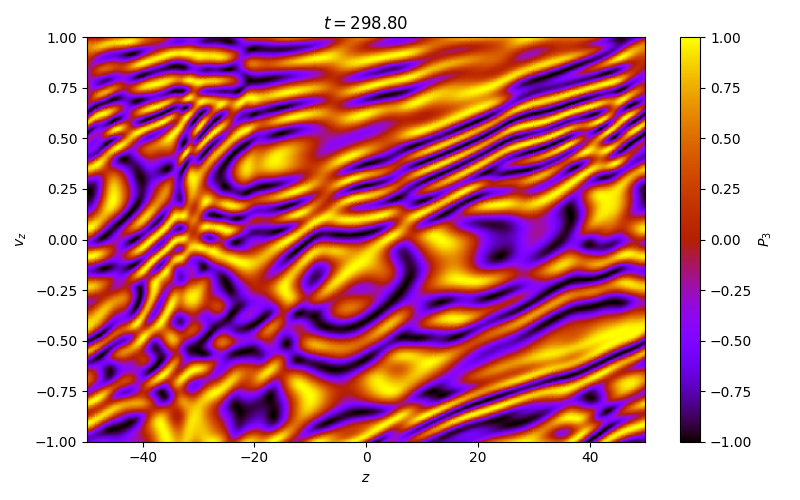}
    \includegraphics[width=0.49\textwidth]{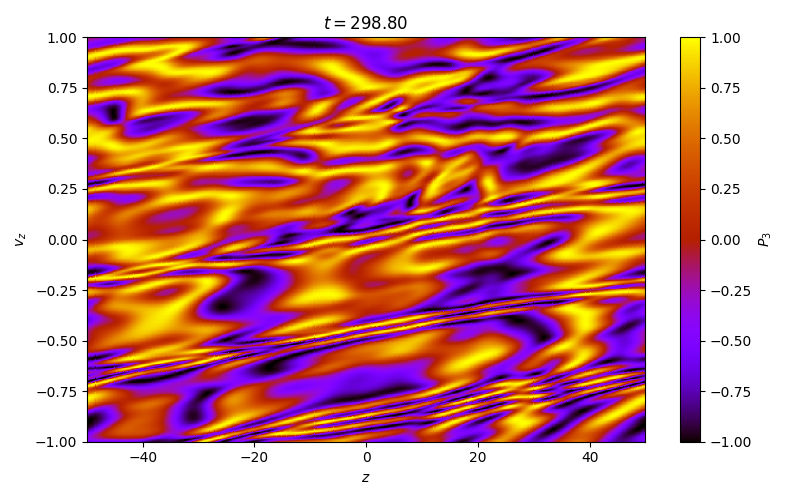}
    \includegraphics[width=0.49\textwidth]{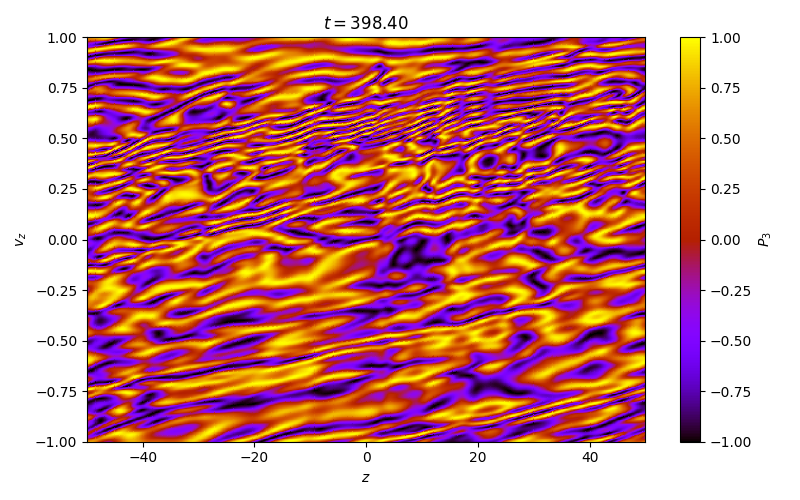}
    \includegraphics[width=0.49\textwidth]{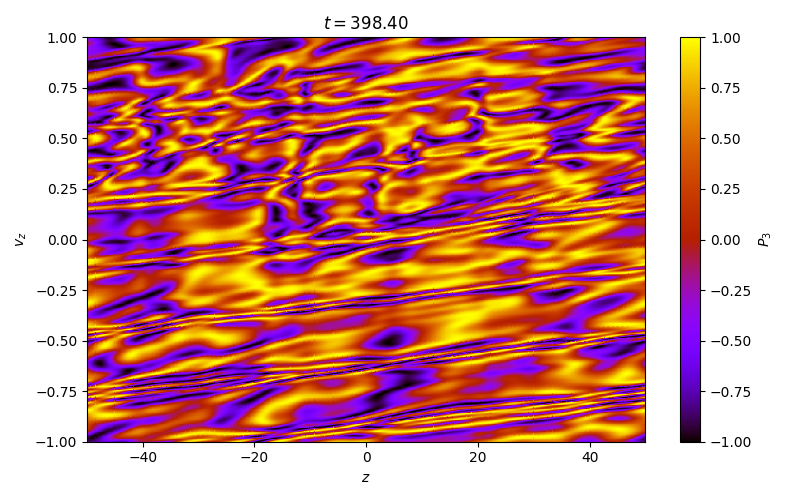}
    \caption{Time evolution of the $P_{3}(t, z_,v_z)$ component of the neutrino polarization vector in the domain box spanned by the spatial variable $-50 \leq z \leq 50$ (horizontal axis) and the angular variable $-1 \leq v_z \leq 1$ (vertical axis). 
    The evolution corresponds to the benchmark system with $\alpha=0.925$ and $\sigma=0.6$. 
    The panels in the left column correspond to snapshots of the flavor evolution for the $\omega=0.01$ (NO) case, while the right panels to the $\omega=-0.01$ (IO) case. 
    }
    \label{fig:P3_on_v_z_plane}
\end{figure*}

\smallskip
\textbf{\textit{Dependence on mass ordering and role of instabilities. ---}}In almost all cases, antineutrinos in the IO tend to evolve toward a state closer to perfect equipartition. While there is no obvious explanation for this difference, we speculate that this behavior is associated with the distinct instability characteristics of the neutrino gas in the linear regime for NO and IO. In particular, as illustrated in Fig.~\ref{fig:lsa_NO_IO}, in IO, a wider range of Fourier modes is unstable in the linear regime (and with higher growth rate), implying that the initial evolution of inhomogeneity-induced decoherence could proceed more efficiently in IO, with the homogeneous mode ($k=0$) being almost the dominantly unstable mode. This fundamental difference between NO and IO is also associated with the different shapes of the eigenvectors of the unstable modes, as illustrated in Fig.~\ref{fig:eigen_vectors}. 
While in the IO case, all angle bins are almost equally affected by instabilities, in the NO case, the impact varies depending on the angle.

\begin{figure}[tb!]
    \centering
    \includegraphics[width=.9625\columnwidth]{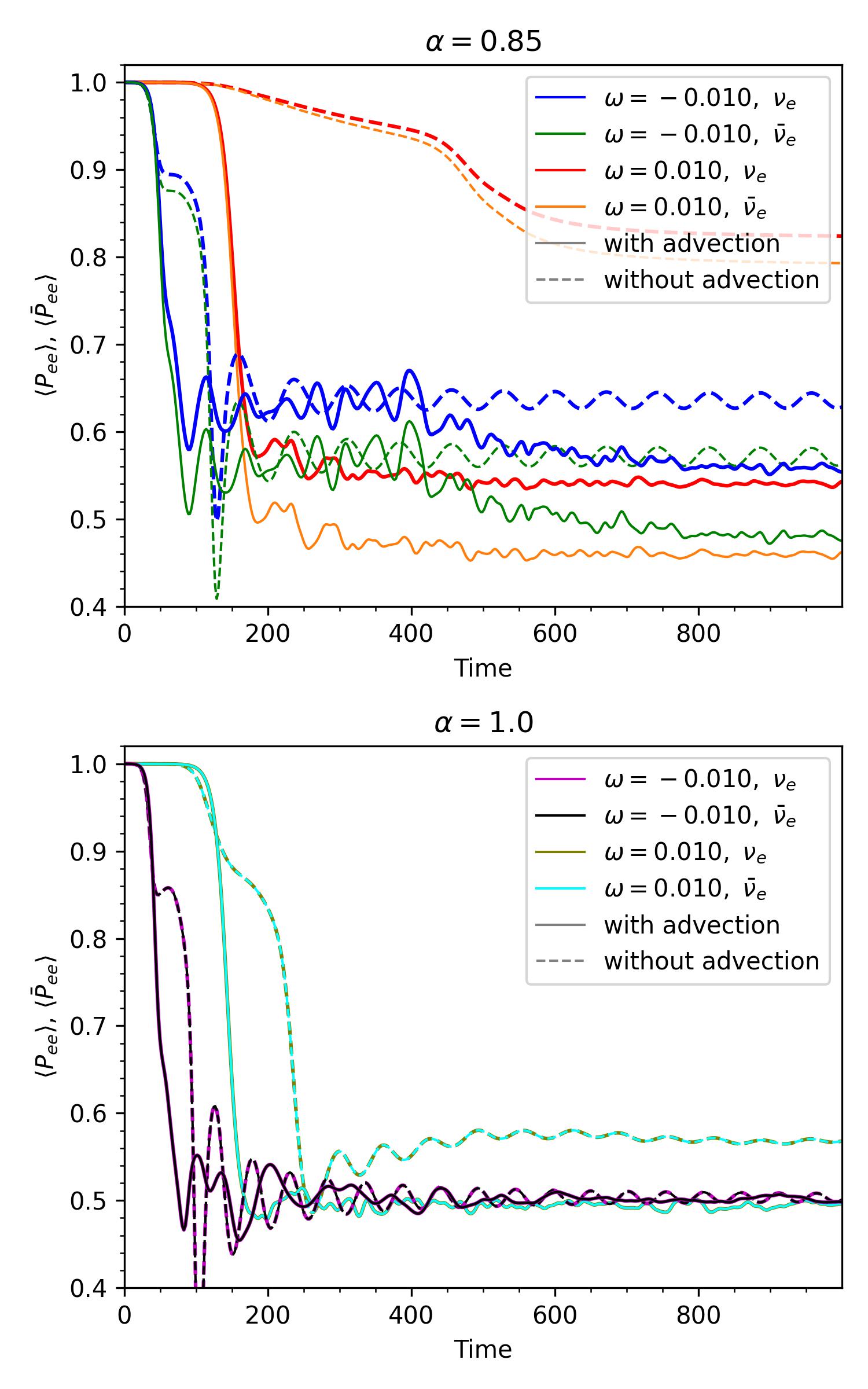}
    \caption{Same as Fig.~\ref{fig:three_alphas}, but showing the results with and without advection. Solid lines show results with advection, while dashed lines represent those without the advection term. Neutrino and antineutrino are shown in corresponding colors. For $\alpha = 0.85$, the case with advection drives the final state of $\langle P_{ee}\rangle$ and $\langle \bar{P}_{ee}\rangle$ closer to 0.5 than in the case without advection. Note that for $\alpha = 1$, flavor equipartition is achieved in the IO case, regardless of whether advection is included.
    }
    \label{fig:adv_compar}
\end{figure}

\begin{figure}[tb!]
    \centering
    \includegraphics[width=.95\columnwidth]{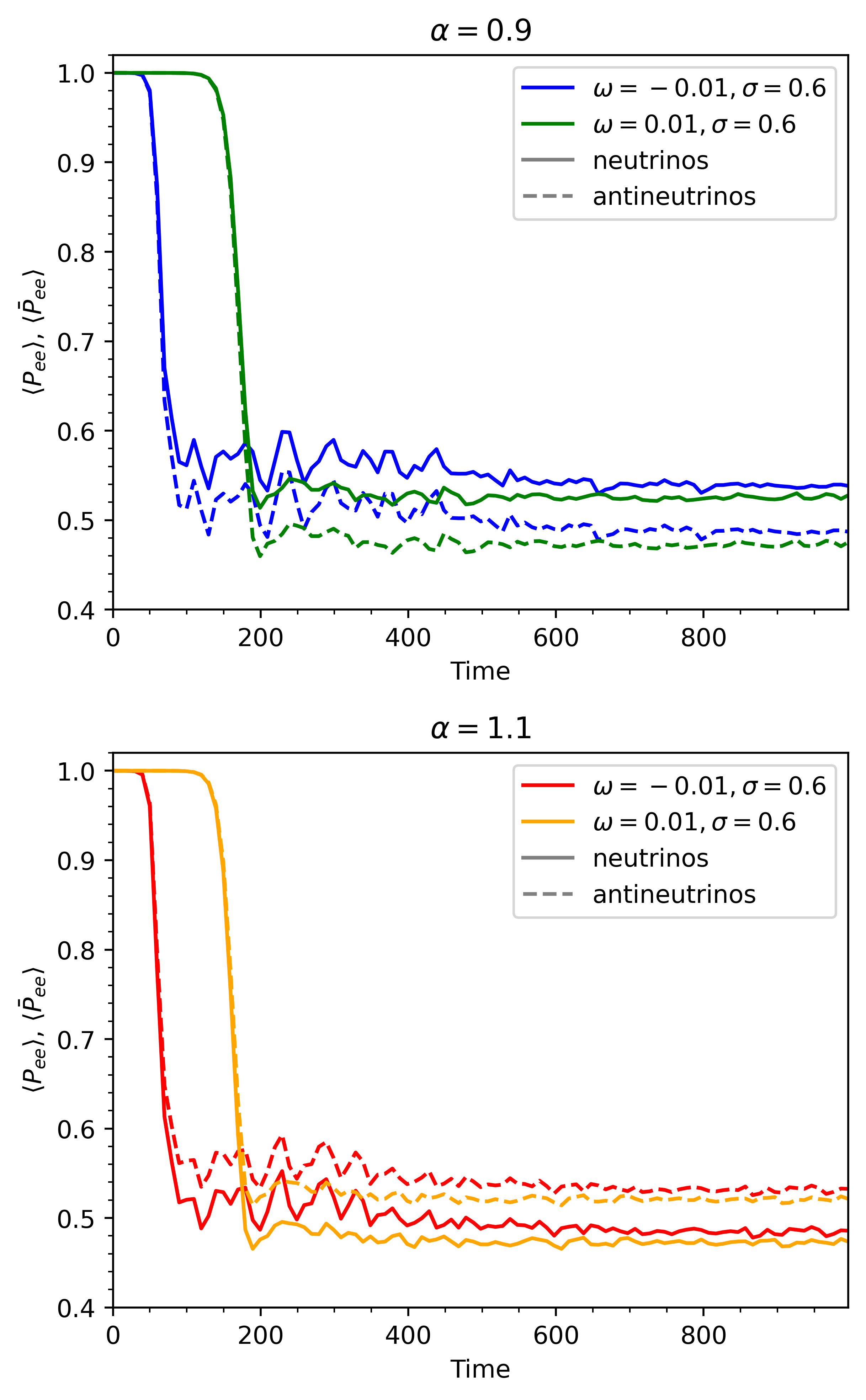}
    \caption{Same as Fig.~\ref{fig:three_alphas}, but with $\alpha = 0.9, 1.1$, while keeping the other parameters fixed, namely $\sigma=0.6$ and $\omega=\pm 0.01$. The roles of the averaged $\langle P_{ee} \rangle$ and $\langle \bar{P}_{ee} \rangle$ are swapped in the $\alpha=0.9$ and $\alpha=1.1$ cases which are symmetric to $\alpha=1$ (see top panel of Fig.~\ref{fig:lsa_colormaps}). 
    The change of sign in the neutrino-neutrino Hamiltonian is responsible for the swap of roles in the flavor dynamics, which is clearly visible in the blue and red curves, where the solid and dashed lines are swapped in both panels, reflecting the change of roles between neutrinos and antineutrinos. 
    The same feature can be seen in the the shape of the angular distributions shown in Fig.~\ref{fig:alpha_grt1_angular}.
    }
    \label{fig:alpha_grt1}
\end{figure}

The characteristics observed in the linear stability analysis can also manifest themselves in the evolution of the neutrino gas in the nonlinear regime. 
This is illustrated clearly in Fig.~\ref{fig:power_spectrum}. 
While the \emph{equilibrated} time-averaged Fourier power spectrum of the off-diagonal component of the neutrino density matrix shows robust and consistent shapes for NO and IO across different profiles (upper panel), the neutrino gas evolves faster toward 
the quasisteady state in IO, as illustrated in the lower panel. 

The impact of instabilities in the linear regime can also be observed in the results in the nonlinear regime presented in Fig.~\ref{fig:P3_on_v_z_plane}. 
As time evolves, the neutrino gas in the  IO goes into the nonlinear regime earlier. Consequently, FCs are activated more rapidly on short scales. 
However, over time, both the NO and IO reach a saturated state, where FCs persist on short scales for both mass orderings. Additionally, the cascade to small angular scales is more efficient in NO than in IO for $t \gtrsim 200$ (see bottom panels of Fig.~\ref{fig:P3_on_v_z_plane}). The emergence of small-scale angular structure can be quantified by computing the first few angular moments of the neutrino distributions. Although not shown here, we have verified that the higher-order moments are larger in NO than in IO, despite IO developing earlier. This suggests distinct dynamical behaviors between the two mass orderings.

\begin{figure}[tb!]
    \centering
    \includegraphics[width=1.\columnwidth]{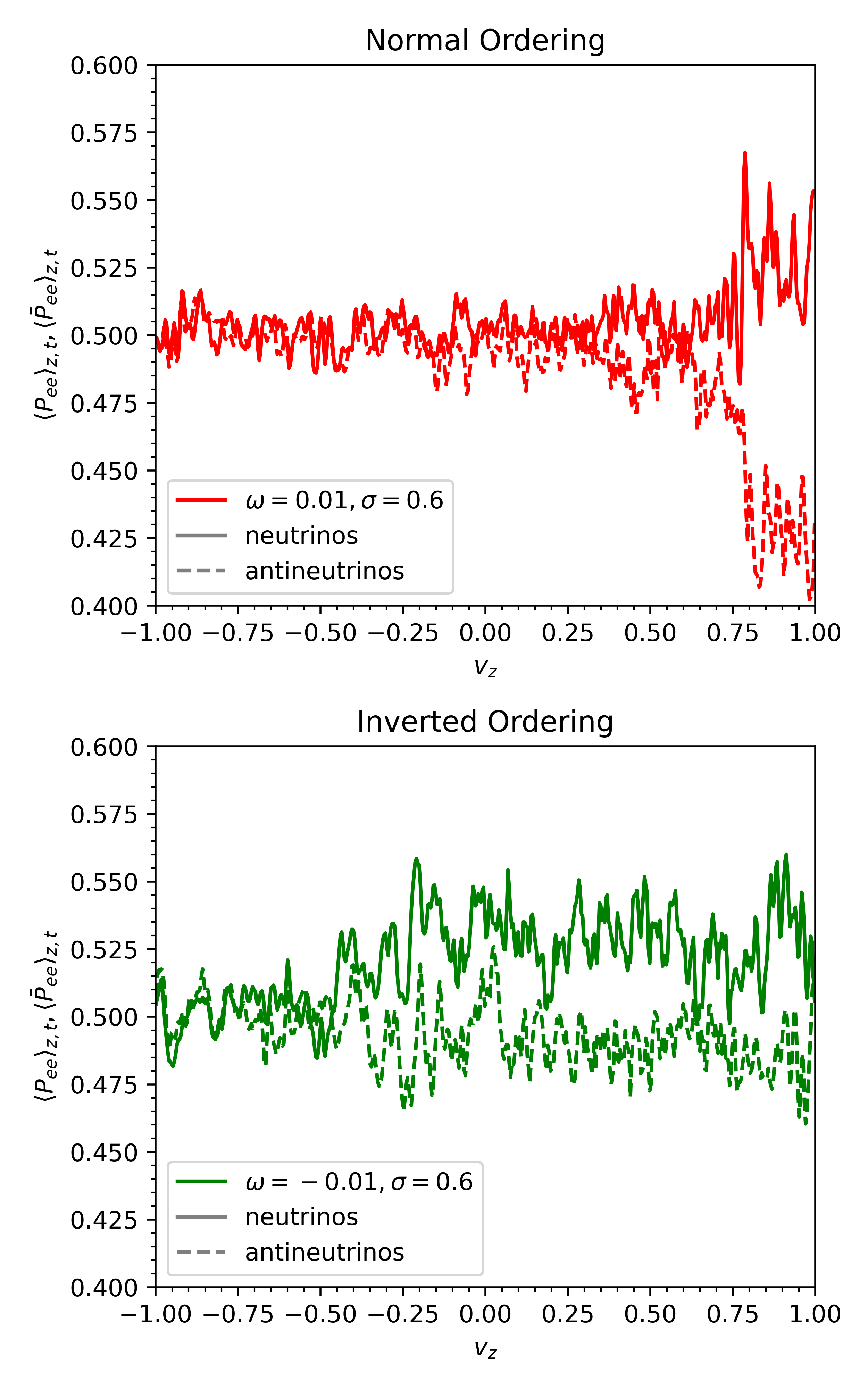}
    \caption{Time-averaged angular distributions of the (spatially-averaged) survival probabilities $\langle P_{ee} \rangle$ and $\langle \bar{P}_{ee} \rangle$ for the $\alpha=0.925$ case shown in Fig.~\ref{fig:three_alphas} and Fig.~\ref{fig:two_sigmas}. The time average is taken over the interval $t=[700, 1000]$, within which the system has reached a quasi-steady state.
    In the NO case, below a certain $v_z\sim 0.6$ value both $\langle P_{ee} \rangle$ and $\langle \bar{P}_{ee} \rangle$ are fully depolarized. For $v_z\gtrsim 0.6$ the departure from flavor equilibration increases. In the IO case, deviation from flavor equilibration is less correlated with $v_z$. 
    }
    \label{fig:two_sigmas_angular}
\end{figure}

\begin{figure}[tbh!]
    \centering
    \includegraphics[width=1.\columnwidth]{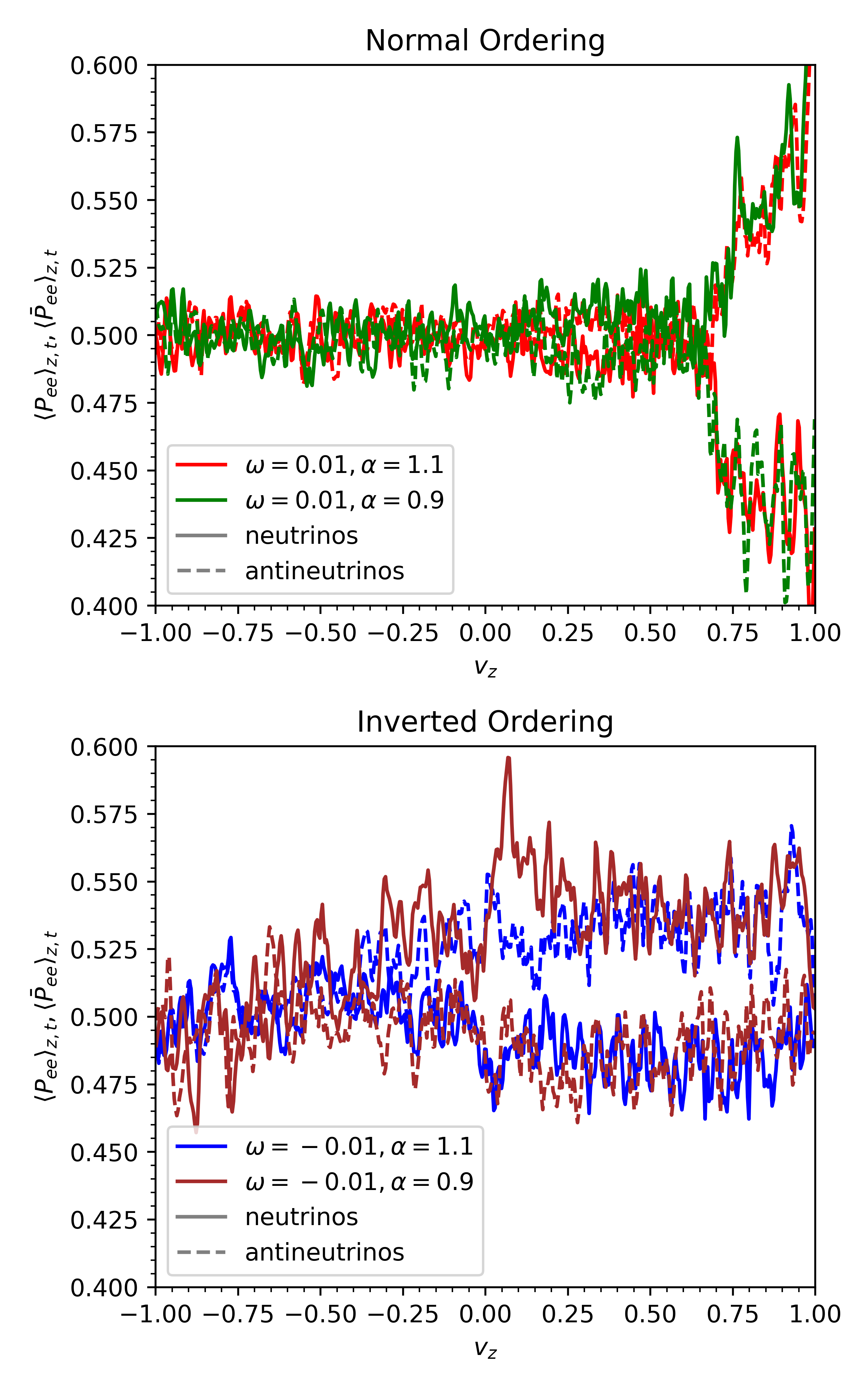}
    \caption{Time-averaged angular distributions of the (anti)neutrino survival probabilities shown in Fig.~\ref{fig:alpha_grt1}, including a case with $\alpha>1$. These results show that flavor equilibration  occurs in the same  manner for both $\alpha\leq1$ and $\alpha>1$, with the only difference being the swapped roles of neutrinos and antineutrinos. 
    }
    \label{fig:alpha_grt1_angular}
\end{figure}

\begin{figure*}[tbh!]
    \centering
    \includegraphics[width=0.49\textwidth]{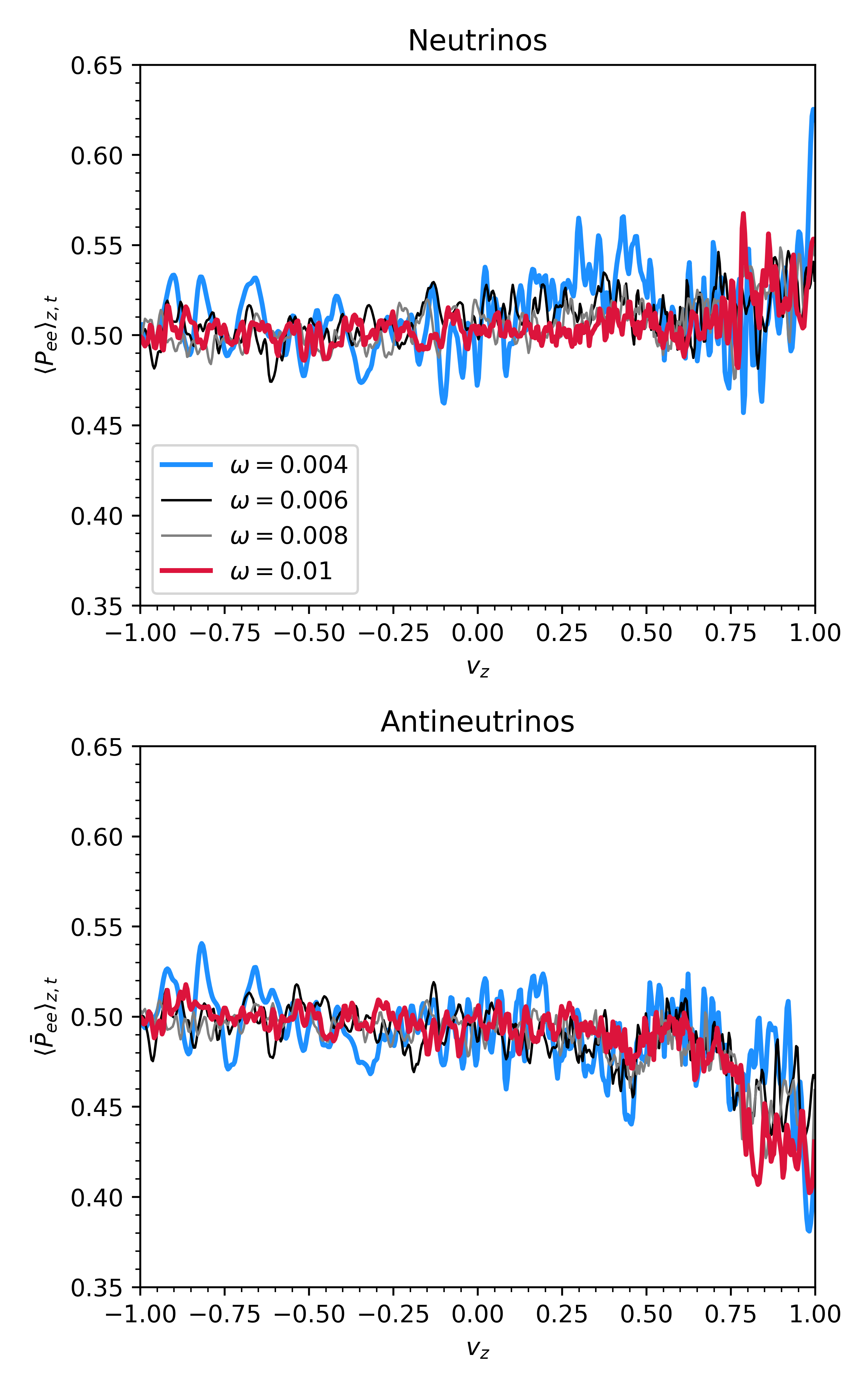}
    \includegraphics[width=0.49\textwidth]{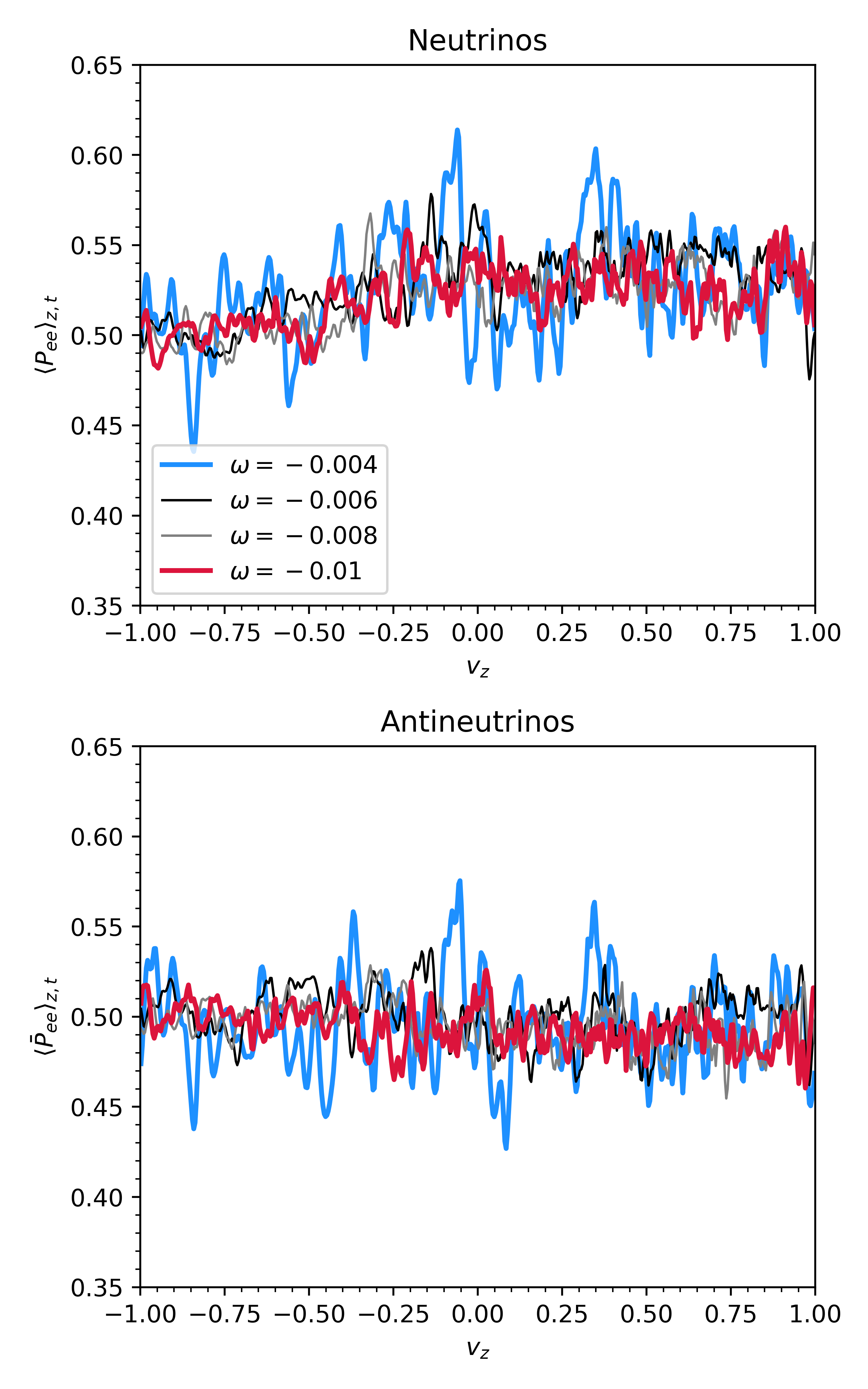}
    \caption{Angular distributions of the time-averaged survival probabilities of systems with varying vacuum frequency within the range $\omega = [0.01, 0.004]$ for NO (left panels) and IO (right panels). We keep the other parameters unchanged, e.g., $\sigma=0.6$ and $\alpha=0.925$. 
    To a zeroth order approximation,
    the overall angular shape of the survival probabilities does not depend strongly on $\omega$ in IO. However, the dip in the forward direction gets suppressed in the case of NO for smaller $\omega$'s. Also note that the $\omega=\pm 0.004$ have notably stronger wiggles in the angular distributions.
    }
    \label{fig:all_omegas_NO_IO}
\end{figure*}

\smallskip
\textbf{\textit{Impact of inhomogeneities and comparison to previous studies.---}}While near flavor equipartition was also observed to occur due to the multi-angle-induced decoherence previously reported in 1D models for $\alpha$ values very close to 1 for IO~\cite{Raffelt:2007yz}, our key finding is that in the presence of inhomogeneities, this becomes a more generic outcome for the neutrino gas. 
As shown in Fig.~\ref{fig:adv_compar}, the neutrino gas reaches near flavor equipartition for $\alpha = 1$ cases without advection in IO (consistent with what was observed in homogeneous models). 
Near equipartition for all other cases occurs only when advection is included in the model. 
This implies that the equilibration unearthed in this study can be considered as an inhomogeneity-induced decoherence rather than a multi-angle-induced decoherence discovered previously in the literature.

Our observation of equilibration is also broadly consistent with the results presented in Ref.~\cite{Martin:2019kgi} for a two-beam neutrino gas~\footnote{Note that although \cite{Martin:2019kgi} evolves the two-beam system in two spatial dimensions, it can be mapped into an equivalent two-beam model that evolves in one temporal and one spatial dimension as the setup in this work.}. 
Although the two-beam setup is computationally much more feasible, it has important limitations for the objectives of this study.
For instance, it does not provide the angular information necessary to calculate the non-zero moments of the neutrino angular distributions. 
Furthermore, we have observed that decoherence in the neutrino gas evolves more readily in the two-beam scenario, leading to equilibration closer to complete equipartition for the subdominant neutrino species. 
Therefore, it is crucial to study equilibration with continuous angular distributions to ensure that the observed equilibration is physically accurate.

\smallskip
\textbf{\textit{Flavor equilibration for $\boldsymbol{\alpha >  1}$ and angular distribution features.---}}So far, we have only discussed cases where $\alpha \leq 1$. 
A very similar flavor equilibration outcome can occur for $\alpha > 1$, as shown in Fig.~\ref{fig:alpha_grt1}. 
However, there is an important yet predictable difference in this scenario: the roles of neutrinos and antineutrinos are reversed. 
Specifically, for $\alpha > 1$, it is the neutrinos that reach near flavor equipartition, while the antineutrinos adjust to a value that ensures the approximate (E-X)LN conservation. 
In addition and similar to $\alpha < 1$, neutrinos reach a state closer to equipartition for IO.

By studying the angle-averaged survival probabilities, one can gain essential information about the flavor content of the neutrino gas. 
However, a significant part of the picture remains incomplete. 
Of particular interest is the equilibration pattern of the neutrino angular distributions. As illustrated in Fig.~\ref{fig:two_sigmas_angular}, the equilibrated angular distributions of the survival probabilities reveal intriguing features.

In the case of IO, the antineutrinos achieve almost a perfect equipartition in momentum space, indicating once again that the (anti)neutrino gas in the IO has a stronger tendency to undergo flavor equipartition (note that $\alpha<1$). 
On the other hand, neutrinos reach a distribution which is peaked mildly in the forward direction, constrained by conservation laws.

In the case of NO, the situation is more complex. 
For antineutrinos, there is a dip in the forward direction, resulting in survival probabilities consistently below 0.5. In contrast, neutrinos exhibit a peak in the forward direction.
The results for the case with $\alpha>1$ are shown in Fig.~\ref{fig:alpha_grt1_angular}.
As noted earlier, the patterns are again quite similar to the $\alpha < 1$ case, except that the roles of neutrinos and antineutrinos are reversed.

The differences between the neutrino and antineutrino angular distributions in NO and IO are again associated with the differences observed in the linear regime, as indicated in Fig.~\ref{fig:eigen_vectors}. 
While in the IO case, the eigenvectors in the linear regime exhibit little to moderate angular dependence, in the NO case, there are strong peaks in the forward and backward directions.

As illustrated in Fig. \ref{fig:all_omegas_NO_IO}, the overall pattern observed in the neutrino angular distributions remains nearly independent of the neutrino vacuum frequency, although flavor equilibration occurs later for smaller values of $\omega$. Moreover, for smaller values of $\omega$, the dip in the forward direction in NO becomes more suppressed, covering a narrower range of angles. If confirmed in more general systems, this would suggest that smaller $\omega/\mu$ ratios lead antineutrinos to exhibit a stronger tendency toward perfect flavor equipartition (even in the NO case) consistent with previous observations.

Although our discussion has focused on the range $0.7 \lesssim \alpha \lesssim 1.3$, we have also explored cases outside this interval. Simulating the flavor evolution of systems with $|1-\alpha| \gtrsim 0.3$ presents significant challenges: such systems require long integration times ($t \gtrsim 10000$) to reach a quasi-steady-state configuration, even when the associated instability growth rates are not necessarily small. Moreover, beyond $t \simeq 7000$, error propagation is no longer under control (particularly in the angular region close to $v_z=1$), leading to the non-conservation of $|P|$, which deviates from 1 by as much as $20\%$.
More sophisticated simulations, incorporating finer spatial and angular grids with much larger box sizes are required to reliably assess the universality of our findings for cases with larger $|1-\alpha|$.

\section{DISCUSSION AND OUTLOOK}\label{sec:conc}

We have investigated the phenomenon of flavor equilibration in the context of SFCs within a periodic box for a two-flavor monochromatic neutrino gas. 
Our findings indicate that a generic equilibration takes place when the densities are averaged over the periodic domain and the angular distributions in the neutrino gas for $0.7 \lesssim \alpha \lesssim 1.3 $. 
This leads to a state where the subdominant neutrino species systematically approach near-complete equipartition with slight flavor overconversion than perfect equipartition. 
The dominant species adjust accordingly to ensure the conservation laws are satisfied.
Although near equipartition occurs in both the IO and NO cases, it is generally observed that the results in IO are closer to equipartition than in NO. 
We have also suggested an empirical formula that can well approximate the coarse-grained flavor conversion probabilities for both $\nu_e$ and $\bar\nu_e$, utilizing the approximate (E-X)LN conservation when the effective vacuum mixing angle is small. Unlike in the case of FFC, there is no clear separation of scales that would allow the disentanglement of advection, collisions, and flavor conversion. Nevertheless, the asymptotic state examined in this study should be interpreted as a quasi-equilibrium state of a local system that approximately conserves lepton number. Our prediction of this quasi-steady state may be useful for schemes that employ relaxation methods to incorporate FC parametrically into hydrodynamical simulations~\cite{Nagakura:2023jfi, Akaho:2025giw, Liu:2025tnf, Johns:2024dbe}.

While number densities of the subdominant neutrino species reach near equipartition for both IO and NO, the neutrino angular distributions exhibit a bit of distinct patterns. 
For $\alpha < 1$, in the case of IO, antineutrinos achieve nearly perfect equipartition in momentum space, whereas neutrinos follow a distribution that is dominated by electron neutrinos in the forward direction (given our gaussian initial angular distributions).
In the case of NO, the situation is more complex. 
Antineutrinos develop a dip in the forward direction, whereas neutrinos exhibit a peak in the forward direction. 
Note that for cases with $\alpha>1$, the main change would be that the roles of neutrinos and antineutrinos are reversed.

Our observations regarding the differences between the neutrino and antineutrino angular distributions in NO and IO may be linked to the distinctions observed in the linear regime. 
In the IO case, the eigenvectors in the linear regime exhibit little to moderate angular dependence, whereas in the NO case, strong peaks appear in the forward and backward directions.

Unlike the multi-angle-induced decoherence previously reported in  1D models for IO, which occurs only for $\alpha$ values very close to 1, the equilibration observed in this work is much more generic. More importantly, it is driven by a different mechanism, namely it is induced by inhomogeneities~\cite{Martin2019a}.

While we observe generic flavor equilibration in the neutrino gas for $0.7 \lesssim \alpha \lesssim 1.3$ which covers most relevant range of the lepton asymmetry ratios in CCSNe except for the neutronization burst phase. For systems with $|1-\alpha| \gtrsim 0.3$, they face significant challenges as they require significantly longer times to reach the quasistationary state. 
Additionally, these systems demand  simulations with finer grids and larger box sizes to mitigate error propagation. We leave such cases for future work.

While our results offer valuable insight into the occurrence of generic flavor equilibration driven by SFC, several important directions remain for future work. In particular, it is essential to extend the model to incorporate realistic neutrino energy spectra and to perform simulations with three neutrino flavors, with the final goal of developing a framework that can be implemented CCSN simulations. Potential utilization of machine learning~\cite{Abbar:2023kta,Abbar:2024chh} or neural networks~\cite{Abbar:2023ltx,Abbar:2024ynh,Richers:2024zit} that may help incorporate SFC into CCSN simulations remain to be explored.
Evolution of SFC under a non-stationary~\cite{Fiorillo:2024qbl,Liu:2024nku} or inhomogeneous background~\cite{Sigl:2021tmj,Bhattacharyya:2025gds}, as well as their interplay with FFC and CFI should all be considered in future studies.

We have here primarily discussed SFCs in the context of CCSNe, but the effect we examined may also be relevant for neutron star mergers, which offer similarly high neutrino densities that are known to host FFC~\cite{Wu:2017qpc,Padilla-Gay:2020uxa,George:2020veu,Li:2021vqj,Just:2022flt,Grohs:2022fyq,Fernandez:2022yyv,Nagakura:2023wbf,Lund:2025jjo,Qiu:2025kgy,Nagakura:2025hss}, CFI~\cite{Xiong:2022zqz,Nagakura:2025hss}, and matter-neutrino resonances~\cite{Malkus:2014iqa,Wu:2015fga,Zhu:2016mwa,Frensel:2016fge,Padilla-Gay:2024wyo}.
Further studies investigating the potential relevance of SFC in mergers are needed to clarify its role.

\textit{\textbf{Acknowledgments.}}--- This work was initiated in part at Aspen Center for Physics, which is supported by National Science Foundation grant PHY-2210452. 
We thank Soumya Bhattacharyya for pointing out connections between earlier literature and this work.
IPG acknowledges support from the U.S. Department of Energy under contract number DE-AC02-76SF00515. 
HHC and MRW acknowledge support of the National Science and Technology Council, Taiwan under Grant No.~111-2628-M-001-003-MY4, the Academia Sinica under Project No.~AS-IV-114-M04. 
MRW also acknowledges support of the Physics Division of the National Center for Theoretical Sciences, Taiwan. 
SA was supported by the German Research Foundation (DFG) through the Collaborative Research Centre ``Neutrinos and Dark Matter in Astro- and Particle Physics (NDM),'' Grant No.\ SFB-1258\,--\,283604770, and under Germany’s Excellence Strategy through the Cluster of Excellence ORIGINS EXC-2094-390783311.
ZX acknowledges support of the European Research Council (ERC) under the European Union’s Horizon 2020 research and innovation program (ERC Advanced Grant KILONOVA No. 885281) and under the ERC Starting Grant (NeuTrAE, No. 101165138).
Part of this work used ASGC (Academia Sinica Grid- computing Center) Distributed Cloud resources, which is supported by Academia Sinica.
We also acknowledge use of the software \textsc{Matplotlib}~\cite{Matplotlib2007}, \textsc{Numpy}~\cite{NumPy2020}, \textsc{SciPy}~\cite{SciPy2020}, and \textsc{IPython}~\cite{IPython2007}.

The work is partially funded by the European Union. Views and opinions expressed are however those of the author(s) only and do not necessarily reflect those of the European Union or the European Research Council Executive Agency. Neither the European Union nor the granting authority can be held responsible for them.

\appendix

\section{Conserved quantities}
\label{appendix:conservedQ}

In this section, we derive the conserved quantities of the system in the absence of spatial inhomogeneities.
We begin with the QKEs for neutrinos written in the language of the polarization vectors $\vec{P}_{\omega,v_z,z}$:
\begin{widetext}
\begin{eqnarray}\label{eq:PolEOMs}
    (\partial_t + v_z \partial_z)\vec{P}_{\omega,v_z,z} &=& \vec{H}_{\omega} \times \vec{P}_{\omega,v_z,z} 
    +\Big[ \int  d\omega^\prime dv_z^\prime g(\omega^\prime, v_z^\prime) \vec{P}_{\omega^\prime,v_z^\prime,z} (1- v_z v_z^\prime) \Big] \times \vec{P}_{\omega,v_z,z} \ .
\end{eqnarray}
\end{widetext}
For each spatial grid the length of the individual polarization vectors $|\vec{P}_{\omega,v_z,z}|$ is conserved during the evolution.
We assume the notation in which 
\begin{eqnarray}
    \vec{H}_{\omega}=\pm \omega \begin{pmatrix} 
    \sin{2 \theta_V} \\
    0 \\
    \cos{2 \theta_V}
    \end{pmatrix} \ ,
\end{eqnarray}
with $+\omega$ for neutrinos and $-\omega$ for antineutrinos.
We assume a two-energy mode system e.g. two possibilities of vacuum frequencies, although our findings carry out straightforwardly to multi-energy calculations. 
Here, we keep $\theta_V$ arbitrarily large not to restrict ourselves to the matter suppressed assumption just yet.

The polarization vector is initially given by
\begin{eqnarray}
    \vec{P}_{\omega,v_z,z}^{t_0} = \begin{pmatrix} 
    0 \\
    0 \\
    1
    \end{pmatrix} \ ,
\end{eqnarray}
for both neutrinos and antineutrinos with the function $g(\omega^\prime, v_z^\prime)$ weighting the integrals such that $g > 0$ for neutrinos and $g < 0$ for antineutrinos. 
We define the angle-integrated polarization  vectors for neutrinos and antineutrinos:
\begin{eqnarray}
    \vec{S}_{\nu} &\equiv& \int dv_z g(\omega, v_z) \vec{P}_{\omega,v_z,z}  h(\omega) \ , \\
    \vec{S}_{\bar{\nu}} &\equiv& \int dv_z g(\omega, v_z) \vec{P}_{\omega,v_z,z}  h(-\omega) \ ,
\end{eqnarray}
where $h(\omega)$ is the Heaviside step function. 
We can now write down the sum and the difference of the angle-integrated polarization vectors given by
\begin{eqnarray}
    \vec{S} &\equiv& \vec{S}_{\nu} + \vec{S}_{\bar{\nu}} \ , \\
    \vec{D} &\equiv& \vec{S}_{\nu} - \vec{S}_{\bar{\nu}} \ ,
\end{eqnarray}
as well as the neutrino (antineutrino) flux vectors
\begin{eqnarray}
    \vec{F}_{\nu} &\equiv& \int dv_z g(\omega, v_z) v \vec{P}_{\omega,v_z,z}  h(\omega) \ , \\
    \vec{F}_{\bar{\nu}} &\equiv& \int dv_z g(\omega, v_z) v \vec{P}_{\omega,v_z,z}  h(-\omega) \ .
\end{eqnarray}
Analogous to the sum and difference of the total polarization vectors, we can define a sum and a difference of fluxes given by
\begin{eqnarray}
    \vec{F}_{+} &\equiv& \vec{F}_{\nu} + \vec{F}_{\bar{\nu}} \ , \\
    \vec{F}_{-} &\equiv& \vec{F}_{\nu} - \vec{F}_{\bar{\nu}} \ .
\end{eqnarray}
Turning back our attention to the QKEs, we can take advantage of these definitions to derive the conserved quantities of the system.  
Integrating the QKEs for neutrinos and antineutrinos (Eq.~\ref{eq:PolEOMs}) over $dv$, one arrives to the following equations in terms of the total polarization vectors:
\begin{eqnarray}
    \partial_t \vec{S}_{\nu} + \partial_z \vec{F}_{\nu} =  &+&\vec{H}_\omega \times \vec{S}_{\nu} \nonumber \\
    &+& \vec{S}_{\bar{\nu}}\times \vec{S}_{\nu} - \vec{F}_{\bar{\nu}}\times \vec{F}_{\nu} \ , \\ 
    \partial_t \vec{S}_{\bar{\nu}} + \partial_z \vec{F}_{\bar{\nu}} =  &-&\vec{H}_\omega \times \vec{S}_{\bar{\nu}} \nonumber \\
    &+& \vec{S}_{\nu}\times \vec{S}_{\bar{\nu}} - \vec{F}_{\nu}\times \vec{F}_{\bar{\nu}} \ .
\end{eqnarray}
Combining these two equations leads to
\begin{eqnarray}
    \partial_t \vec{S} + \partial_z \vec{F}_{+} &=& \vec{H}_{\omega} \times \vec{D} \ , \\
    \partial_t \vec{D} + \partial_z \vec{F}_{-} &=& \vec{H}_{\omega} \times \vec{S} + \vec{S} \times \vec{D} + \vec{F}_{+}\times \vec{F}_{-} \ .
\end{eqnarray}
Taking the first equation and computing the inner product with the vacuum vector, we get
\begin{eqnarray}
    \partial_t (\vec{S}\cdot \vec{H}_{\omega}) + \partial_z (\vec{F}_{+}\cdot \vec{H}_{\omega}) = 0 \ .
\end{eqnarray}
Moreover, performing the spatial average simplifies the above equation, e.g. $\partial_z \langle \vec{F}_{+}\cdot \vec{H}_{\omega} \rangle = 0$, leading to the conserved quantity of the system:
\begin{eqnarray}
    \partial_t \langle \vec{S}\cdot \vec{H}_{\omega} \rangle = 0 \ . 
\end{eqnarray}
During the evolution of the system, the parallel component $\langle S_{||}\rangle \equiv \langle \vec{S}\cdot \vec{H}_{\omega}\rangle $ remains a constant of motion. 
In the special case where the vacuum mixing angle is matter-suppressed $\theta_V \ll 1$, the direction of $\vec{H}_{\omega}$ almost aligns with the flavor direction $\hat{z}$. 
This imposes an approximate conservation of (E-X)LN that relates the equilibrium value of $\langle P_{ee} \rangle$ to that of $\langle \bar{P}_{ee} \rangle$, as described by Eq.~\ref{eq:avePee} in the main text. 
Throughout this work, the component $\langle S_{||}\rangle$ is conserved to a good approximation.


\bibliography{Library}


\onecolumngrid

\clearpage

\end{document}